\documentclass[fleqn,usenatbib]{mnras}
\usepackage{mathptmx}
\usepackage[T1]{fontenc}
\DeclareRobustCommand{\VAN}[3]{#2}
\let\VANthebibliography\thebibliography
\def\thebibliography{\DeclareRobustCommand{\VAN}[3]{##3}\VANthebibliography}

\usepackage{graphicx}
\graphicspath{{./images/}}
\usepackage{amsmath}
\usepackage{cleveref}
\crefformat{equation}{Equation~#2#1#3}
\usepackage{upgreek}
\usepackage{tipa}
\usepackage{mathtools}
\usepackage{xcolor}
\usepackage{bm}
\usepackage{verbatim}

\usepackage{amsmath}	
\usepackage{amssymb}	

\DeclareMathOperator{\sech}{sech}
\newcommand{\eagle}{\textsc{eagle}}

\def \aap{A\&A}

\def \apj{ApJ}
\def \apjl{ApJ Lett.}
\def \apjs{ApJS}

\def \araa{ARA\&A}

\def \mnras{MNRAS}

\def \prl{Phys. Rev. Lett.}

\DeclareMathOperator\erf{erf}

\newcounter{example}[section]

\title[Disc heating \& galaxy morphology]{The impact of spurious collisional heating on the morphological evolution of simulated galactic discs}

\author[Wilkinson et al.] {\parbox{17.6cm}{
    Matthew J. Wilkinson$^{1,2}$\thanks{E-mail: matthew.wilkinson@research.uwa.edu.au},
    Aaron D. Ludlow$^{1,2}$,
    Claudia del P. Lagos$^{1,2}$,
    S. Michael Fall$^{3}$,
    Joop Schaye$^{4}$ and
    Danail Obreschkow$^{1,2}$
  }\vspace{0.3cm}\\
  $^{1}${International Centre for Radio Astronomy Research, University of Western Australia, 35 Stirling Highway, Crawley, Western Australia, 6009, Australia}\\
  $^{2}${
ARC Centre of Excellence for All Sky Astrophysics in 3 Dimensions (ASTRO 3D).
  }\\
  $^{3}${Space Telescope Science Institute, 3700 San Martin Drive, Baltimore, MD 21218, USA.
  }\\
  $^{4}${Leiden Observatory, Leiden University, PO Box 9513, 2300 RA Leiden, the Netherlands}\\
}

\date{Accepted XXX. Received YYY; in original form ZZZ}

\pubyear{2022}

\begin{document}
\label{firstpage}
\pagerange{\pageref{firstpage}--\pageref{lastpage}}
\maketitle

\begin{abstract}
  We use a suite of idealised N-body simulations to study the impact of spurious heating of star particles by dark matter particles on the kinematics
  and morphology of simulated galactic discs. We find that spurious collisional heating leads to a systematic increase of the azimuthal velocity
  dispersion ($\sigma_\phi$) of stellar particles and a corresponding decrease in their mean azimuthal velocities ($\overline{v}_\phi$).
  The rate of heating is dictated primarily by the number of dark matter halo particles (or equivalently, by the dark matter particle mass at fixed
  halo mass) and by radial gradients in the local dark matter density along the disc; it is largely insensitive to the stellar particle mass.
  Galaxies within haloes resolved with fewer than $\approx 10^6$ dark matter particles are particularly susceptible to spurious morphological
  evolution, irrespective of the total halo mass (with even more particles required to prevent heating of the galactic centre). Collisional heating
  transforms galactic discs from flattened structures into rounder spheroidal systems, causing them to
  lose rotational support in the process. It also affects the locations of galaxies in standard scaling relations that link their various
  properties: at fixed stellar mass, it increases the sizes of galaxies, and reduces their mean stellar rotation velocities and specific
  angular momenta. Our results urge caution when extrapolating simulated galaxy scaling relations to low masses where spurious collisional
  effects can bias their normalisation, slope and scatter.
\end{abstract}



\section{Introduction}

Galaxies are complex dynamical systems, and much of our understanding of their origin and evolution has been inferred from simulations. 
In particular, cosmological simulations -- which are now capable of sampling the diverse environments in which galaxies form, while
simultaneously resolving their internal properties -- yield galaxy populations that appear realistic when confronted
with wide array of observational data \citep[e.g.][]{Schaye2015, Furlong2017, Ludlow2017, Lagos2017, Nelson2018, vandeSande2019, Pillepich2019}. 
Recent advances in algorithms and computing architecture have also led to increases in both simulation volume and particle number, and
large-volume cosmological simulations (i.e. those with volumes of order $\approx 100$ Mpc cubed or greater) routinely achieve mass
and force resolutions of order $10^6\,{\rm M_\odot}$ and $1~{\rm kpc}$, respectively \citep{Vogelsberger2014, Schaye2015, Pillepich2018b, Dave2019}.
Despite these improvements, even the most massive galaxies and dark matter haloes formed in such simulations are still resolved with far fewer
particles than there are stars and dark matter (DM) particles in real galaxies, and are therefore subject to incoherent fluctuations in
the gravitational potential of point-like particles. 
These fluctuations deflect the trajectories of stellar and DM particles (through a scattering process commonly referred to as ``collisions''),
causing them to deviate from the smooth paths dictated by the mean-field potential of the system.

Galaxies are dynamically cold stellar systems
embedded within comparatively hot DM haloes, and collisions between their constituent particles may therefore alter the structure and kinematics
of galaxies in undesirable ways. For example, collisions can lead to a net exchange of energy between the two components as they attempt to reach
energy equipartition (the galaxy heats up in the process, while the halo cools down). This effect becomes more pronounced as the mass ratio of DM
to stellar particles, $\mu\equiv m_{\rm DM}/m_\star$, is increased, but is also present when $\mu=1$ due to the strong phase-space segregation of
stars and DM typical of most galaxies. But for disc galaxies, which harbour most of their kinetic energy in ordered rotation, collisions between DM
and stellar particles also transfer kinetic energy from azimuthal stellar motions into kinetic energy in radial and vertical motions, even when the 
energy exchanged between the two components is small. As a result, disc galaxies are particularly susceptible to spurious evolution due to collisional
effects.

\citet[][]{Lacey1985} calculated analytically the collisional heating rate of thin galactic discs embedded within
DM haloes composed of point-mass particles (assumed to be black holes, which were considered plausible DM candidates
at the time). Their calculations are based on epicyclic theory and assume that the disc remains thin and cold relative to the halo at
all times, i.e. $\sigma_\star\ll\sigma_{\rm DM}$. By comparing their analytic results to the observed velocity dispersion of
stars in the solar neighbourhood, they concluded that black holes with masses exceeding $\gtrsim 10^6\, {\rm M_\odot}$ cannot dominate
the DM in the Milky Way's halo or else collisional heating would have rendered its disc hotter and thicker than it is observed to be.
This critical black hole mass is of order (or smaller than) the mass of DM particles used in many large-volume cosmological
simulations, suggesting many simulated galaxies are affected by spurious collisional heating. Indeed,
\citet[][see also \citealt{Revaz2018}]{Ludlow2019} showed that ${\rm M_\star}\lesssim 10^{10}{\rm M_\odot}$ galaxies in the \eagle~
simulation \citep[][which has a DM particle mass of $m_{\rm DM}\approx 10^7\,{\rm M_\odot}$]{Schaye2015} undergo spurious size growth
due to collisional heating. 

Later, \citet{Ludlow2021} used idealised\footnote{This refers to non-cosmological simulations, typically of isolated, equilibrium
  systems. Our idealised runs consist of a two-component model composed of an initially thin disc and a
spherically-symmetric DM halo (see Section~\ref{sec:sims} for details).} simulations of
secularly-evolving, stable galactic discs to confirm the theory developed
by \citet[][]{Lacey1985}, and proposed empirical corrections to their analytic results able to accommodate instances of extreme disc heating
(i.e. when $\sigma_\star\approx \sigma_{\rm DM}$ and the vertical scale height of the disc is of order its scale length, which can occur
in poorly-resolved haloes). Using their empirical
model, they verified that the stellar discs of Milky Way-mass galaxies, if simulated using DM particles with masses
$m_{\rm DM}\gtrsim 10^6\,{\rm M_\odot}$, are subject to non-negligible levels of spurious heating. For example, their vertical
velocity dispersion will artificially increase by $\gtrsim 20 \,{\rm km/s}$ and their scale heights by $\gtrsim 300\, {\rm pc}$
in roughly a Hubble time. These values are of the same order of magnitude as the observed vertical velocity dispersion and scale
height of Milky Way disc stars.

Spurious collisional heating by dark matter particles affects the kinematics of disc stars, their rotation velocities, as well as the
radial and vertical sizes of discs.\footnote{Previous work on the spurious collisional heating of stellar discs
      focused primarily on the self-scattering of disc particles \citep[e.g.][]{Sellwood1987,RLWhite1988,Sellwood2013}. This differs
      from our work, which instead emphasises the effects of spurious heating due to collisions between stellar particles and more
      massive dark matter halo particles.}
And because cosmological simulations typically sample the DM density field using equal mass particles,
low-mass galaxies are resolved with fewer particles than massive ones and are therefore more vulnerable to the effect. This
may lead to mass-dependent biases in the standard
scaling laws that link the structural and kinematic properties of simulated galaxies, such as their masses, sizes and
characteristic velocities, as well as how these relations depend on galaxy morphology.

From an observational perspective, the slopes, intercepts, and scatter of many galaxy scaling relations are well-constrained,
with errors of order $10$ per cent or less \citep[e.g.][]{Di-Teodoro2023}. Reproducing these scaling laws is a primary goal of galaxy
formation models, and it is therefore necessary to draw comparisons between observed and simulated galaxies only when the
latter are free from numerical artefacts. 

This paper is a follow-up to \citet{Ludlow2021} in which we address these issues. It is organised as follows.
In Section~\ref{sec:sims} we describe our simulations (Section~\ref{ssec:sims}) and
analysis techniques (Section~\ref{ssec:analysis}). Section~\ref{sec:heating} contains our main results:
We begin with an overview of the morphological evolution of galactic discs due to
spurious collisional heating (Section~\ref{ssec:visual}), followed by quantitative analyses of its effect on their azimuthal
velocity profiles (Section~\ref{ssec:azimuthal_vel}), galaxy scaling relations (Section~\ref{ssec:scaling-relations}), and morphologies
(Section~\ref{ssec:measurements}). In Section~\ref{sec:model} we present a model that describes the evolution of the velocity dispersions
and azimuthal velocities of stellar disc particles (Section~\ref{ssec:model}) and discuss the implications of our results for
cosmological simulations (Section~\ref{ssec:cosmosims}). We provide our conclusions in Section~\ref{sec:summary}.

\section{Simulations and Analysis}
\label{sec:sims}

\subsection{Simulations}
\label{ssec:sims}

Our analysis is based on the suite of idealised disc galaxy simulations first presented in \citet{Ludlow2021}, which we
review below. The initial conditions\footnote{The initial conditions for all of our simulations were created using \texttt{GalIC}
  \citep[see][for details]{Yurin2014}.} of our simulations comprise an initially thin, axisymmetric stellar disc of mass ${\rm M}_\star$ embedded
within a spherically-symmetric and isotropic dark matter (DM) halo. We adopt a coordinate system coincident with the galaxy centre
and align the $z$-axis with the disc's angular momentum vector, $\mathbfit{J}_\star$. In this coordinate system, the three-dimensional structure of
the disc is described by 
\begin{equation}
    \rho_\star(R,z)=\frac{{\rm M_\star}}{4\pi\, z_d\, R_d^2}\exp\biggr(-\frac{R}{R_d}\biggl)\,\sech^2\biggr(\frac{z}{z_d}\biggl),
    \label{eq:star-density}
\end{equation}
where $z$ denotes the vertical height above the disc plane and $R=\sqrt{x^2+y^2}$ is the distance from the
$z$-axis; $z_d$ and $R_d$ are the scale height and scale length of the disc, respectively.

The DM halo is modelled as a
\cite{Hernquist1990} sphere of total mass ${\rm M_{DM}}$, which has a circular velocity profile given by 
\begin{equation}
    V_{\rm DM}(r)=\sqrt{\frac{{\rm G\,M_{DM}}\, r}{(r+a)^2}},
    \label{eq:Vchern}   
\end{equation}
where $a$ is its characteristic scale radius, $G$ is the gravitational constant, and $r=\sqrt{R^2+z^2}$ is the three-dimensional
radial coordinate. Where necessary, we distinguish the circular velocity due to DM and stars using subscripts, and use
$V_c$ to denote the total circular profile due to the halo and disc: $V_c^2(r)=V_{\rm DM}^2(r) + V_\star^2(r)$.

Equation~(\ref{eq:Vchern}) predicts a circular velocity profile similar to that of a Navarro-Frenk-White profile
\citep[][hereafter, NFW]{Navarro1996,Navarro1997} at $r\lesssim a$, which allows us to characterise the halo's
structure in terms of the more familiar virial velocity,\footnote{Throughout the paper, we define the virial radius of a DM
  halo as that of a sphere enclosing a mean density of $\rho_{200}\equiv200\,\rho_{\rm crit}$, where $\rho_{\rm crit}=3\,H^2/8\,\pi\,G$ is
  the critical density for closure and $H$ is the Hubble-Lema$\hat{i}$tre constant. This
  implicitly defines the halo's virial mass, ${\rm M}_{200}$, and virial circular velocity, $V_{200}=\sqrt{G\,{\rm M}_{200}/r_{200}}$.}
$V_{200}$, and concentration, $c$, of the latter \citep[e.g.][]{Springel2005b}. We do so by matching their inner density profiles, which
requires a Hernquist scale radius of $a=(r_{200}/c)\sqrt{2\,f(c)}$, where $f(c)=\ln(1+c)-c/(1+c)$, and a total halo mass of
${\rm M_{DM}}=(1-f_\star)\,{\rm M_{200}}$, where $f_\star$ is the stellar mass fraction.
Other structural parameters of relevance are the halo's spin parameter, $\lambda_{\rm DM} = j_{\rm DM} / (\sqrt{2} \, r_{200} V_{200})$,
and the ratio of the specific angular momentum of the disc to that of the halo, i.e. $f_j=j_\star/j_{\rm DM}$ (which is sometimes referred
to as the angular momentum retention fraction). 

Much of our analysis is based on a suite of ``fiducial'' galaxy models for which we adopt
$V_{200} = 200\, {\rm km/s}$, $c = 10$ (or equivalently, $r_{200}/a\approx 5.8$),
$\lambda_{\rm DM} = 0.03$, and $f_j=1$ (i.e. we assume that the disc and halo initially have the same specific angular momentum,
which is approximately valid for both observed and simulated disc galaxies; e.g. \citealt{Di-Teodoro2023}; \citealt{Rodriguez-Gomez2022}).
The latter determines the initial size of the disc, which for our fiducial models is $R_{d} = 0.02 \, r_{200} \approx 4.1\,{\rm kpc}$.
The initial disc scale height, which is independent of $R$, is chosen to be $z_{d} = 0.05 \, R_d$ ($\approx 0.2 \,{\rm kpc}$
for our fiducial models). All of our simulations adopt a stellar mass fraction of $f_\star=0.01$. 
We note that this is lower than the inferred stellar-to-halo mass ratio of discs occupying haloes with virial velocities
of $V_{200} \approx 200\, {\rm km/s}$, which is closer to $0.05$ \citep[e.g.][]{Posti2019, Di-Teodoro2023}. However, it ensures that our
isolated, equilibrium discs are: 1) free from Toomre instabilities that may jeopardise the
collisional heating effects we wish to quantify; and 2) not massive enough to gravitationally alter the structure of the surrounding DM
halo, which would complicate the interpretation of our results (see Appendix A of \citealt{Ludlow2021} for details).

Our suite of fiducial models differ only in their (stellar and DM) mass resolution. For the DM, we adopt a range
of particle masses corresponding to integer and half-integer values of $\log m_{\textrm{DM}}$, starting from a lowest-resolution
of $m_{\rm DM}=10^8 \, {\rm M}_\odot$, and extending to $m_{\rm DM}=10^6 \, {\rm M}_\odot$ (the corresponding number of DM halo
particles span $N_{\rm DM}= 1.8 \times  10^4$ to $1.8 \times 10^6$).
The stellar particle mass is determined by the DM-to-stellar particle mass ratio, for which we adopt $\mu\equiv m_{\rm DM}/m_\star = 5$.
This is approximately the inverse of the cosmic baryon fraction \citep[i.e. $\Omega_{\rm DM}/\Omega_{\rm bar}\approx 5.36$; e.g.][]{Planck2018},
and roughly equivalent to the (initial) $\mu$ value for cosmological smoothed particle hydrodynamics simulations that adopt equal
numbers of DM and baryonic particles. 

In addition to our fiducial runs we also analyse models with $V_{200} = 50, 100$ and $400\,\,{\rm km/s}$.
These models also span a range of DM particle
masses that differ by increments of $\Delta\log m_{\rm DM}=0.5$. Specifically, $4\leq \log (m_{\rm DM}/{\rm M_\odot})\leq 6$ for
$V_{200}=50\, {\rm km/s}$, $5\leq \log (m_{\rm DM}/{\rm M_\odot})\leq 7$ for $V_{200}=100\, {\rm km/s}$, and
$7\leq \log (m_{\rm DM}/{\rm M_\odot})\leq 9$ for $V_{200}=400\, {\rm km/s}$. These choices ensure that the
DM haloes are resolved with between $N_{\rm DM}\approx 10^4$ and $\approx 10^6$ DM particles regardless of $V_{200}$. Each of
these additional models adopt the
same values of $\lambda_{\rm DM}$, $f_\star$, $f_j$, $\mu$, and $z_d/R_d$ as our fiducial runs, which ensures that
the structural properties of the discs and haloes of all models scale proportionally. 
Note also that, for all $V_{200}$, the local DM density is the same at radii $R_f$ enclosing a fraction $f$ of the disc's
stellar particles. 

Because our fiducial runs adopt a stellar-to-DM particle mass ratio that differs from unity, the cumulative effects of collisions
results in a transfer of energy from the high- to low-mass particle species (i.e. from DM to stellar particles in our case), albeit
at a rate slower than heating due to shot noise in the DM particle distribution \cite[see Appendix A of][for details]{Ludlow2021}.
This effect is known as ``mass segregation'' and has been shown to result in spurious
size growth of simulated galaxies \citep[e.g.][]{Revaz2018, Ludlow2019}. To assess the impact of mass segregation on
disc morphology, we also consider a suite of models for which $\mu=1$ and 25, although defer a discussion of these
to Appendix \ref{sec:fitting}. 

Collisional heating is sensitive to local scattering events (in addition to distant encounters), and therefore depends on the local
density and characteristic velocities
DM particles (for centrally concentrated systems like galactic haloes), in addition to their masses.
For that reason, we also consider a subset of our $V_{200}=200\,{\rm km/s}$ models for which we adopt different halo concentration
parameters, specifically $c=7$ and $c=15$. For these we adjust $f_j$ such that the stellar mass
profiles remain unchanged. These runs, which are discussed in Appendix \ref{sec:fitting}, were carried out for
$m_{\rm DM}=10^7 \, \rm M_\odot$ and $10^8 \, \rm M_\odot$, and all adopt $\mu=5$.

All of our simulations were performed using \texttt{GADGET-2} \citep{Springel2005} for a total integration time of 9.8 Gyr.
We used a Plummer-equivalent gravitational softening length of $\epsilon_{\rm soft} = z_d$, which marginally resolves the
vertical forces across the disc. As discussed in \citet{Ludlow2021}, the impact of spurious collisional heating on
  the properties of our simulated discs is independent of gravitational softening provided $\epsilon_{\rm soft}\lesssim z_d$;
  heating is suppressed for larger softening values, but not eliminated.\footnote{This reflects the fact that gravitational
    scattering occurs for both short- and long-range interactions, with the latter unaffected by softening
    \citep[e.g.][]{Huang1993,Theis1998,Dehnen2001,Ludlow2019b}.}
  In this paper we only consider models with $\epsilon_{\rm soft}=z_d$, as these ensure that gravitational
  forces are marginally resolved across the galaxy disc and that our results are not strongly influenced by
  softening. We note, however, that our $\epsilon_{\rm soft}$ values are likely smaller than those adopted for 
  the majority of cosmological hydrodynamical simulations.

For most simulations, we output 100 snapshots equally-spaced
in time $t$; for those corresponding to 
haloes with fewer than $N_{\rm DM} < 1.2\times 10^5$ particles we increase the output cadence by a factor of 4
(i.e. we output 400 snapshots in total).\footnote{As mentioned below, many of the graphical results presented in this paper
    were smoothed over a time interval of $\Delta t=2\,{\rm Gyr}$ for aesthetic purposes. The increased output frequency adopted for
    our low-resolution runs improves the robustness of the smoothing operation without affecting our results.}
We note that the initial structure of the disc and halo in our poorly-resolved models
is subject to Poisson noise due to the coarse sampling of their respective distribution functions. For that reason,
we carried out 5 (10) simulations based on independent realisations of the initial conditions for models with
$N_{\rm DM} < 1.2\times 10^5$ ($<4\times 10^4$) and present their results as averages over all realisations.
A summary of all models studied in the paper is provided in Table~\ref{table:simulation-list}.

\begin{table*}
  \caption{Properties of the disc galaxies analysed in this paper. The first five columns list the properties of the
    disc or halo that we vary: $V_{200}$ is the virial circular velocity of a
    Navarro-Frenk-White profile with the same inner density structure as our adopted Hernquist dark matter halo;
    $c=r_{200}/r_{-2}$ is the NFW halo's concentration; $f_j=j_\star/j_{\rm DM}$ is the disc's angular momentum
    retention fraction (i.e. the ratio of the disc to halo specific angular momentum); $R_d$ is the scale length of
    the disc, expressed here in units of NFW halo scale radius, $r_{-2}$; and $z_d$ is the disc scale height, expressed in units of
    $R_d$. For each $V_{200}$ we simulate a range of models with different numbers of DM particles, $N_{\rm DM}$ (corresponding to
    different DM particle masses, $m_{\rm DM}$), some of which also vary the the DM-to-stellar particle mass ratio, $\mu=m_{\rm DM}/m_\star$;
    The relevant values are listed under $N_{\rm DM}$, $m_{\rm DM}$ and $\mu$, respectively. These parameters, together with
    the stellar mass fraction, which is $f_\star=0.01$ for all runs, specify the stellar particle masses.
    For our low-resolution models, Poisson noise due to coarse sampling of the disc and halo distribution functions
    leads to slight differences in their initial structure and kinematics. In those cases, we carried our multiple runs corresponding to different
    realisations of the initial conditions and presented results after averaging over all of them. The number of runs
    carrier out for each $N_{\rm DM}$ is listed in the right-most column.}
  \centering
  \begin{tabular}{r r r r r c c c c} 
    \hline \hline
    $V_{200}$ [km/s] & 
    $c$ & 
    $f_j$ & 
    $R_d / r_{-2}$ & 
    $z_d / R_d$ & 
    $N_{\rm DM} / 10^5$ & 
    $\log (m_{\rm DM}/$M$_\odot)$ & 
    $\mu = m_{\rm DM} / m_\star$ &
    Number of runs \\
    \hline
    200  & 10 & 1.0  & 0.20 & 0.05 & 0.18, 0.58, 1.84, 5.82, 18.4 & 8.0, 7.5, 7.0, 6.5, 6.0 & 5 & 10, 5, 1, 1, 1 \\ 
    200  & 7  & 0.81 & 0.14 & 0.05 & 0.18, 1.84                   & 8.0, 7.0                & 5     & 10, 1  \\
    200  & 15 & 1.03 & 0.31 & 0.05 & 0.18, 1.84                   & 8.0, 7.0                & 5     & 10, 1 \\
    200  & 10 & 1.0  & 0.20 & 0.10 & 0.18, 1.84                   & 8.0, 7.0                & 5     & 10, 1 \\
    200  & 10 & 1.0  & 0.20 & 0.20 & 0.18, 1.84                   & 8.0, 7.0                & 5     & 10, 1 \\ 
    200  & 10 & 1.0  & 0.20 & 0.05 & 0.18, 0.58, 1.84, 5.82, 18.4 & 8.0, 7.5, 7.0, 6.5, 6.0 & 1, 25 & 10, 5, 1, 1, 1 \\ 
    50   & 10 & 1.0  & 0.20 & 0.05 & 0.29, 0.91, 2.88, 9.09, 28.8 & 6.0, 5.5, 5.0, 4.5, 4.0 & 5     & 10, 5, 1, 1, 1 \\ 
    100  & 10 & 1.0  & 0.20 & 0.05 & 0.23, 0.73, 2.3, 7.27, 23.0  & 7.0, 6.5, 6.0, 5.5, 5.0 & 5     & 10, 5, 1, 1, 1 \\ 
    400  & 10 & 1.0  & 0.20 & 0.05 & 0.15, 0.47, 1.47, 4.66, 14.7 & 9.0, 8.5, 8.0, 7.5, 7.0 & 5     & 10, 5, 1, 1, 1 \\ 
    \hline
  \end{tabular}
  \label{table:simulation-list}
\end{table*}

\subsection{Analysis}
\label{ssec:analysis}

We use a number of diagnostics to quantify the morphology of galactic discs, some based on the spatial distribution of stellar
particles and others based on their kinematics. We review them below.

We characterise disc size using the cylindrical half-mass radius, $R_{1/2}$, enclosing half of all stellar particles; the analogous
vertical half-mass height is denoted $z_{1/2}$. These are initially related to the scale length and height of the disc via 
$R_{1/2}= 1.68 \, R_d$ and $z_{1/2}=0.55\, z_d$, respectively. Many of the results presented in \Cref{sec:heating} and \Cref{sec:model}
are based on measurements made within a cylindrical shell centred on $R_{1/2}$ of width $\Delta\log R=0.2$,
which is sufficiently wide to enclose $\gtrsim 100$ star particles for all of our models.
However, we have verified that similar results are obtained at other radii $R_f$ enclosing different fractions $f$ of the
disc's stellar mass.

Stellar particle velocities are also evaluated in cylindrical coordinates, and for our analysis we measure their
dispersions in the vertical ($\sigma_z$; i.e. perpendicular to the disc plane), the radial ($\sigma_R$;
i.e. perpendicular to the $z$-axis) and the azimuthal directions ($\sigma_\phi$), although our analysis principally focuses
on the latter. For each velocity component $i$,
the velocity variance is defined $\sigma_i^2=\sum_j\, m_j(v_{i,j}-\overline{v}_i)^2 / \sum_j m_j$, where $\overline{v}_i$ is the
mass-weighted mean velocity in the $i$ direction (i.e. $\overline{v}_i=\sum_j m_j v_{i,j} / \sum_j m_j$), $m_j$ is the mass of
the $j^{\rm th}$ stellar particle, and the sum extends over all particles $j$ that occupy a given cylindrical shell. We
note that our disc galaxies have $\overline{v}_z\approx \overline{v}_R\approx 0$ at all times, and $\overline{v}_\phi\approx V_c$ initially. 

We quantify galaxy morphology using a few diagnostics. The first, $\kappa_{\rm rot}$,
is defined as the fraction of the disc's kinetic energy contained in rotational motion
\citep[e.g.][]{Sales2010}, and is defined by
\begin{equation}
  \kappa_{\rm rot}=\frac{\sum_j m_j v_{\phi,j}^2}{\sum_j m_j v_j^2},
  \label{eq:kappa-data}
\end{equation}
where $v_j$ magnitude of the $j^{\rm th}$ particle's velocity, and the sum extends over all particles
within the annulus. We also consider the stellar spin parameter, $\lambda_r$ \citep[e.g.][]{Emsellem2007, Naab2014}, which is defined as
\begin{equation}
  \lambda_r=\frac{\overline{v}_\phi}{\sqrt{\overline{v}_\phi^2+\sigma^2_{1\rm D}}},
  \label{eq:lambda-data}
\end{equation}
where $\sigma^2_{1\rm D}\equiv (\sigma_z^2+\sigma_R^2+\sigma_\phi^2) / 3$ is the one dimensional stellar velocity
dispersion (squared) at $R$.

Finally, we measure the ratio of rotation-to-dispersion velocities, i.e. $\overline{v}_\phi/\sigma_{1\textrm D}$,
and the circularity of stellar particle orbits, $\varepsilon_{\rm circ}\equiv j_z/j_c(E)$ \citep{Abadi2003}. In the latter, $j_z$ is the
$z$-component of the particle's angular momentum and $j_c(E)$ is the angular momentum of a circular orbit in the plane of the disc
with the same total energy. The circularity parameter can be used to calculate the disc-to-total ($D/T$) mass ratio, which is commonly
defined as the mass fraction of orbits with $\varepsilon_{\rm circ} \geq 0.7$ \citep{Sales2010, Aumer2013, Grand2017, Joshi2020}, i.e.
$D/T=M_\star^{-1}\sum_j^{\varepsilon_{\rm circ}\geq 0.7} m_j$ (we note, however, that this threshold will also include orbits with
grossly non-circular motions, as emphasised by \citealt{Peebles2020}). Similarly, the spheroid-to-total
ratio, $S/T$, is defined as twice the mass fraction of counter-rotating orbits, i.e. $S/T=(2/M_\star)\sum_j^{v_\phi<0} m_j$.
In general, these definitions imply that $D/T + S/T \neq 1$. We calculate $S/T$ and $D/T$ using only particles within a given
cylindrical shell.

In addition to these dynamical measures of galaxy morphology, we also consider several geometric quantities:
1) the disc aspect ratio $z_{1/2}/R_{1/2}$ (note that both quantities are calculated using {\em all} stellar particles); and
2) the ratio $c/a$ of the minor-to-major axis lengths of the galaxy's moment of inertia tensor. In practice, we follow \citet{Thob2019}
and use an iterative scheme to obtain the principle axis lengths of the reduced moment of inertia tensor, which is appropriate for
highly flattened geometries. Note that $c/a$ is calculated using all stellar particles. 

\begin{figure*}
  \includegraphics[width=1\textwidth]{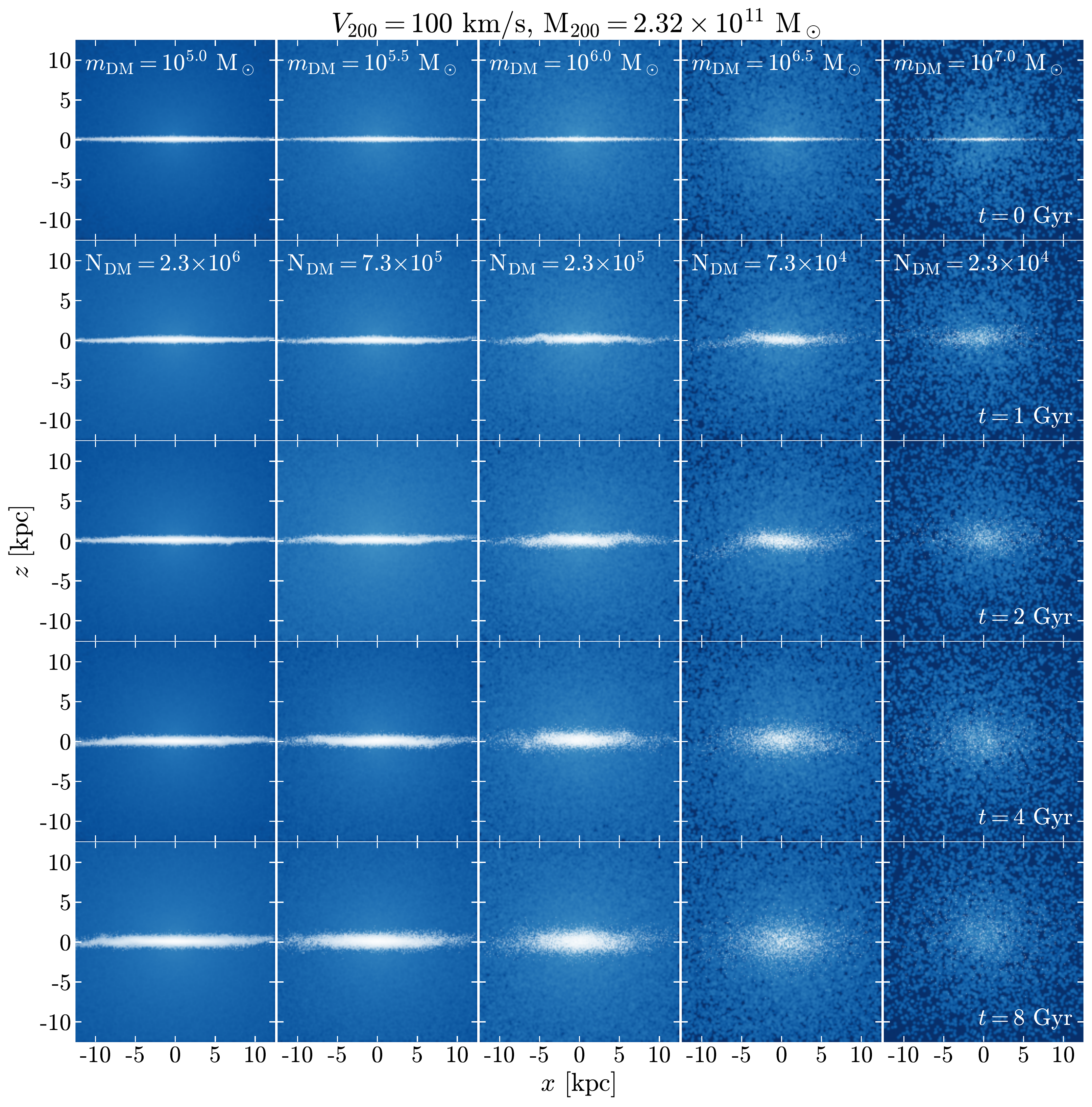}
  \caption{
    Edge-on view of the evolution of a simulated galactic disc consisting of a \citet{Hernquist1990}
    dark matter halo (with $V_{200}=100 \,{\rm km/s}$; blue points) and exponential stellar disc (white points;
    equation~\ref{eq:star-density}). Different rows correspond to different
    output times; from top-to-bottom, $t=0, \, 2, \, 4$ and $8\,{\rm Gyr}$, respectively. 
    From left to right, the different columns correspond to models with DM particle masses,
    $m_\textrm{DM} = 10^5, 10^{5.5}, 10^6, 10^{6.5} \textrm{ and } 10^7 {\rm M}_\odot$, respectively. The
    corresponding particle numbers are
    $N_{\rm DM} = 2.3\times 10^6,\, 7.3\times 10^6,\, 2.3\times 10^5,\, 7.3\times 10^5$, and $2.3\times 10^4$,
    respectively; these values are provided in the various panels.
    The stellar disc thickens over time at a rate that increases with increasing DM particle mass and decreasing
    particle number, and become progressively more spheroidal with time, particularly if the resolution is low.}
  \label{fig:projections-low}
\end{figure*}

\begin{figure*}
  \includegraphics[width=1\textwidth]{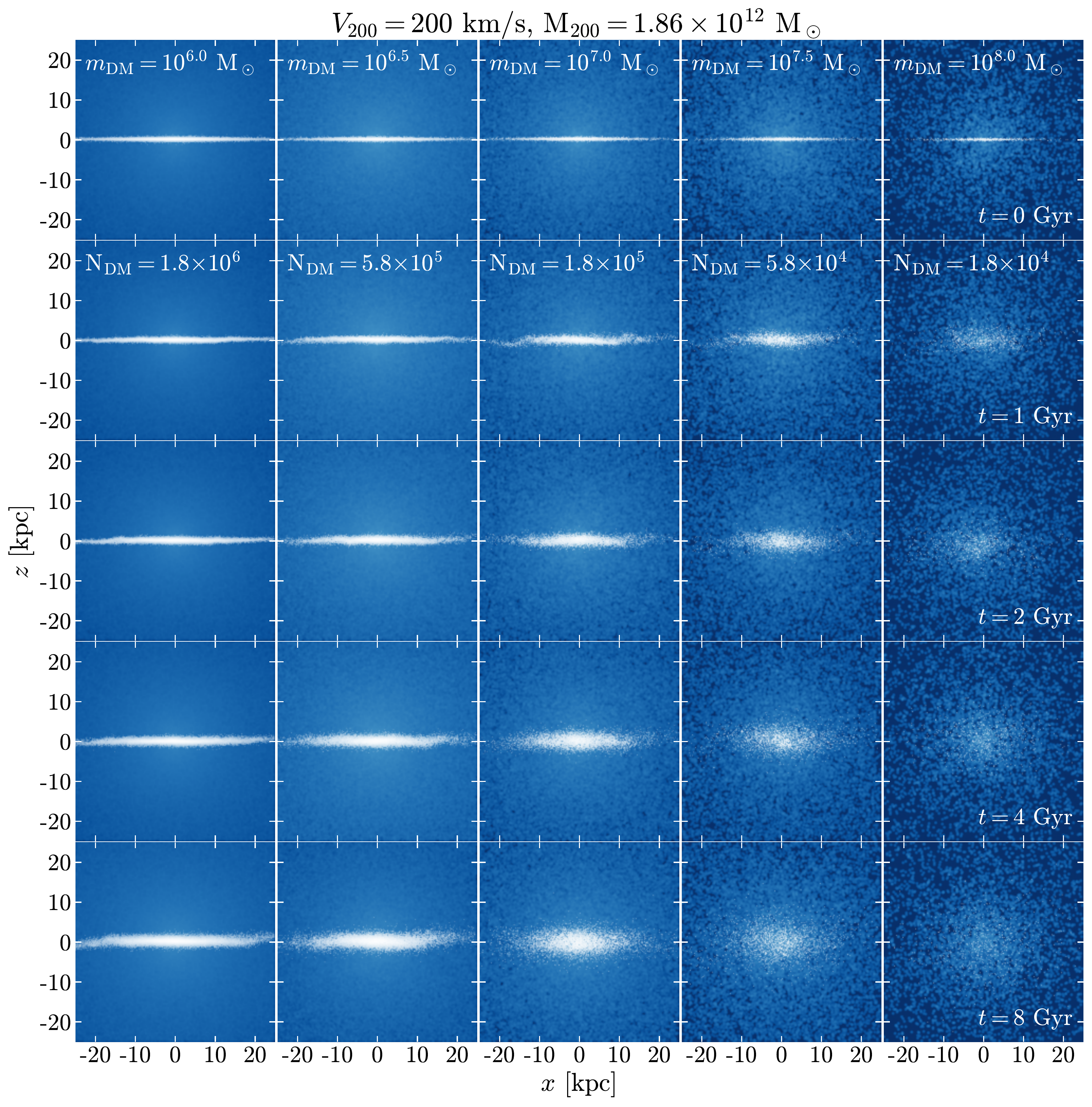}
  \caption{Same as Fig.~\protect\ref{fig:projections-low} but for our fiducial galaxy model, i.e. $V_{200}=200 \,{\rm km/s}$
    (${\rm M}_{200}=1.86\times 10^{12}{\rm M_\odot}$). Note that the mass resolution, $m_{\rm DM}$, of these models differ from
    those used for the corresponding panels of Fig.~\ref{fig:projections-low}, but the values of $N_{\rm DM}$ are similar.}
  \label{fig:projections-med}
\end{figure*}

\begin{figure*}
  \includegraphics[width=1\textwidth]{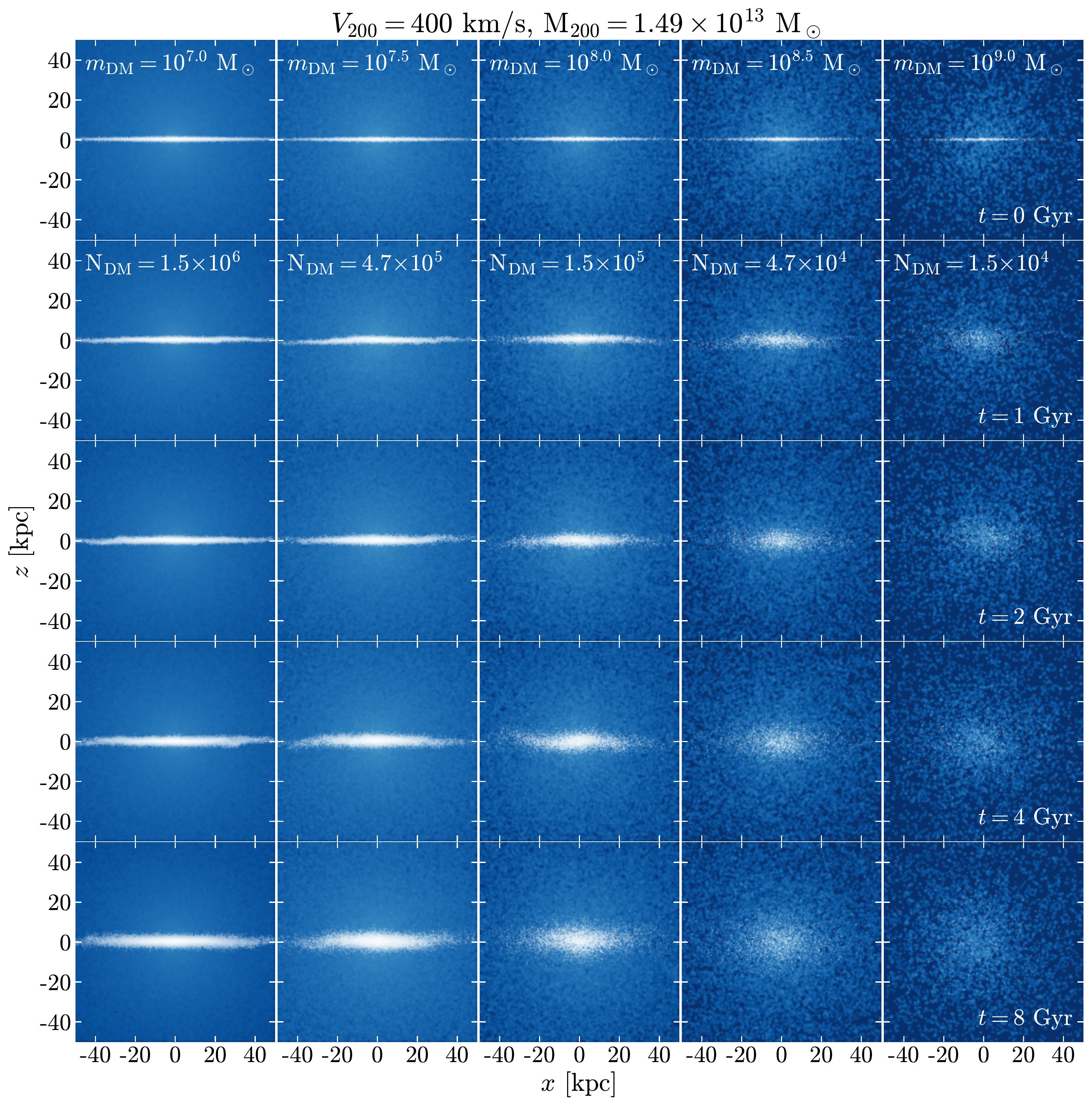}
  \caption{Same as Fig.~\protect\ref{fig:projections-low} but for $V_{200}=400\,{\rm km/s}$
    (${\rm M_{200}=1.49\times 10^{13}\,M_\odot}$). Note the different particle masses used in the corresponding
    columns of Fig.~\protect\ref{fig:projections-low}.}
  \label{fig:projections-hi}
\end{figure*}

\section{The effects of collisional heating on the properties of simulated galactic discs}
\label{sec:heating}

\subsection{Visual appearance}
\label{ssec:visual}

Fig.~\ref{fig:projections-low} shows edge-on projections of five
galaxy models (corresponding $V_{200}=100\, {\rm km/s}$) that are
identical in all respects other than their mass resolution. Stellar and DM particles are represented using
white and blue points, respectively, and have been smoothed for visual presentation using Py-SPHViewer
\citep{Benitez-Llambay2015}. Each disc has an initial half-mass height of $z_{1/2}=57\,{\rm pc}$,
and an initial aspect ratio $z_{1/2}/R_{1/2}=0.016$.

From left-to-right, the DM particle mass increases from
$m_{\rm DM}=10^5\,{\rm M_\odot}$ to $m_{\rm DM}=10^7\,{\rm M_\odot}$ in equally-spaced steps of $\Delta \log m_{\rm DM}=0.5$
(corresponding to haloes resolved with between $N_{\rm DM}=2.3\times 10^6$  and $2.3\times 10^4$ DM particles). From top-to-bottom,
the different rows correspond to different output times, from $t=0$ (i.e. the initial conditions of the simulations)
to $t=1, \, 2, \, 4,$ and $8\,{\rm Gyr}$, respectively. Although these galaxies were constructed to be
in stable equilibrium, they clearly follow different evolutionary paths depending on $m_{\rm DM}$. For
example, for $m_{\rm DM}=10^5\,{\rm M_\odot}$ ($N_{\rm DM}=2.3\times 10^6$; i.e. the best-resolved system, which is plotted in
the left-most panels of Fig.~\ref{fig:projections-low}), the galactic disc remains thin at all times. Its vertical half-mass
height, for example, only increased by a factor of $\approx 2$ over $8\,{\rm Gyr}$ (at which point $z_{1/2}/R_{1/2}\approx 0.037$).
Contrast this with the {\it least}-resolved
galaxy, which is plotted in the right-most panel and has $m_{\rm DM}=10^7\,{\rm M_\odot}$ ($N_{\rm DM}=2.3\times 10^4$). This disc
galaxy is virtually unrecognisable after only $t\approx  2\,{\rm Gyr}$, by which time its
half-mass height has increased by a factor of $\approx 13$ relative to the initial value (and $z_{1/2}/R_{1/2}\approx 0.19$).
But its evolution does not end there:
after $t=8\,{\rm Gyr}$, $z_{1/2}\approx 2\,{\rm kpc}$, a factor of $\approx 33$ larger than at $t=0$. By comparing the
various models at fixed $t$ (i.e. along rows of Fig.~\ref{fig:projections-low}), it is clear that spurious
collisional heating, as observed at fixed times, increasingly affects galaxies in less-resolved haloes.

Figs.~\ref{fig:projections-med} and \ref{fig:projections-hi} show similar results, but for haloes of virial mass
${\rm M}_{200}={\rm 1.86\times 10^{12}\,M_\odot}$ ($V_{200}=200\,{\rm km/s}$; which is of order the mass of the Milky Way) and
${\rm M}_{200}=1.49\times 10^{13}\,{\rm M_\odot}$ $(V_{200}=400\,{\rm km/s}$), respectively. As in Fig.~\ref{fig:projections-low},
the DM particle masses increase from left-to-right, but span a different range from the lowest- to the
highest-resolution cases depending on ${\rm M_{200}}$. As mentioned in \Cref{ssec:sims}, this ensures that the various models plotted
in corresponding columns of Figs.~\ref{fig:projections-low} to \ref{fig:projections-hi} are resolved with similar
{\it numbers} of DM particles, rather than with similar DM particle masses. The visual effects of collisional heating
are nonetheless analogous in all three figures, and clearly show that it is the {\em number} of particles per halo (rather than the particle
masses) that dictates whether or not collisional heating will be an important driver of
morphological evolution. Consider, for example, runs carried out with $m_{\rm DM}=10^7\,{\rm M_\odot}$ (shown in the right-most,
middle and left-most columns of Figs.~\ref{fig:projections-low}, \ref{fig:projections-med}, and \ref{fig:projections-hi},
respectively). For $V_{200}=100\, {\rm km/s}$ (Fig.~\ref{fig:projections-low}) this particle mass corresponds to the lowest-resolution 
simulation (i.e. $N_{\rm DM}=2.3\times 10^4$); but for $V_{200}=400\, {\rm km/s}$ (Fig.~\ref{fig:projections-hi}) it
is the highest-resolution run ($N_{\rm DM}=1.5\times 10^6$). These galaxies evolve very differently.
Both are initially thin, and have half-mass heights of $z_{1/2}=0.016 \, R_{1/2}$ (corresponding to $57\,{\rm pc}$ and
$0.226 \,{\rm kpc}$ for $V_{200}=100\, {\rm km/s}$ and $V_{200}=400\, {\rm km/s}$, respectively). But after $8\,{\rm Gyr}$, $z_{1/2}$
has increased to $z_{1/2} = 0.40 \, R_{1/2}$ ($1.96\,{\rm kpc}$) and $0.047 \, R_{1/2}$ ($0.65\,{\rm kpc}$) 
for $V_{200}=100\,{\rm km/s}$ and $400\,{\rm km/s}$ respectively. This suggests that galaxies of similar ages in
cosmological simulations that inhabit haloes spanning a wide range of masses will be subject to different levels of
spurious heating, a result that is expected but worth re-emphasising.

\subsection{Azimuthal velocity profiles}
\label{ssec:azimuthal_vel}

\begin{figure*}
\includegraphics[width=0.45\textwidth]{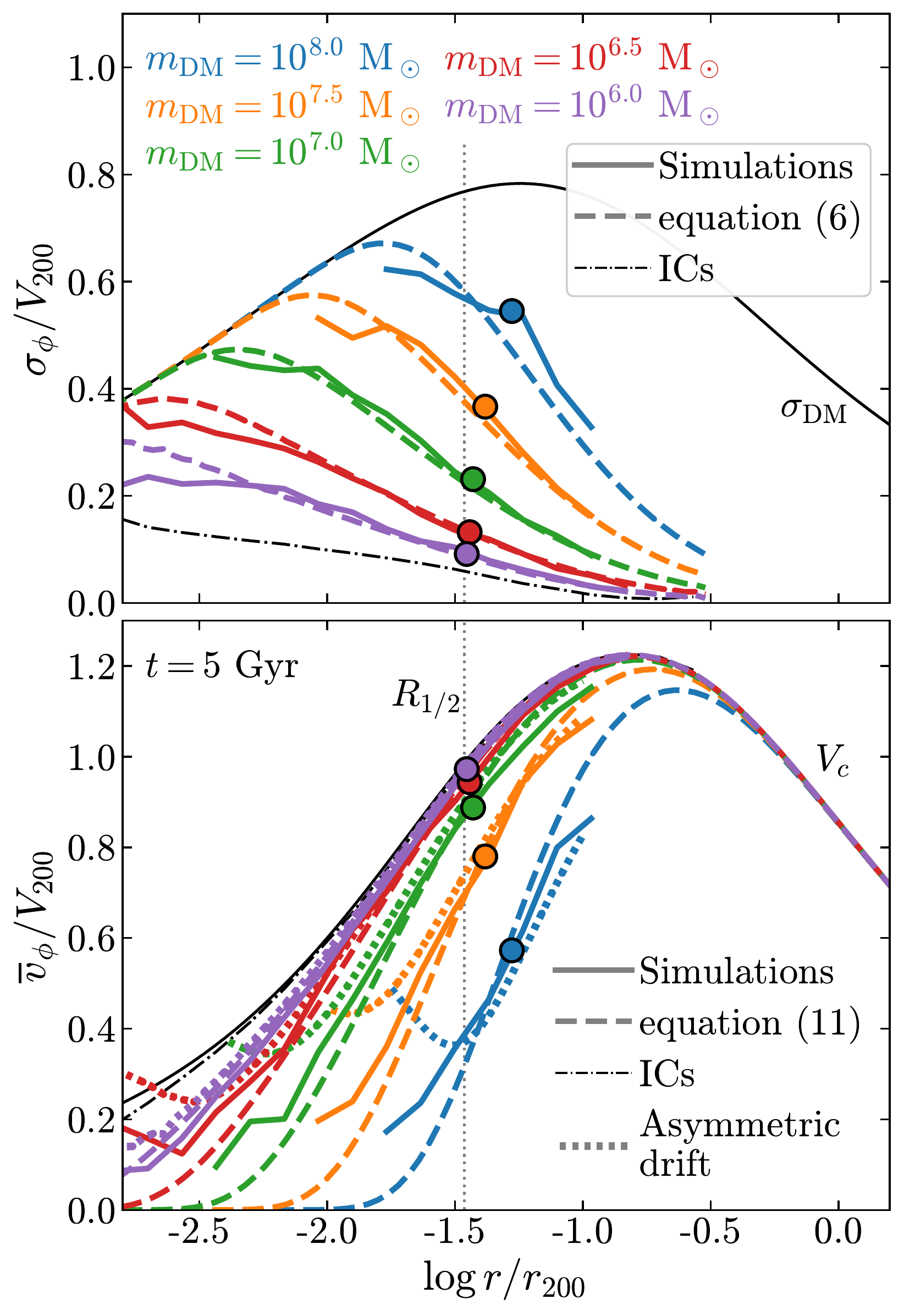}
\includegraphics[width=0.45\textwidth]{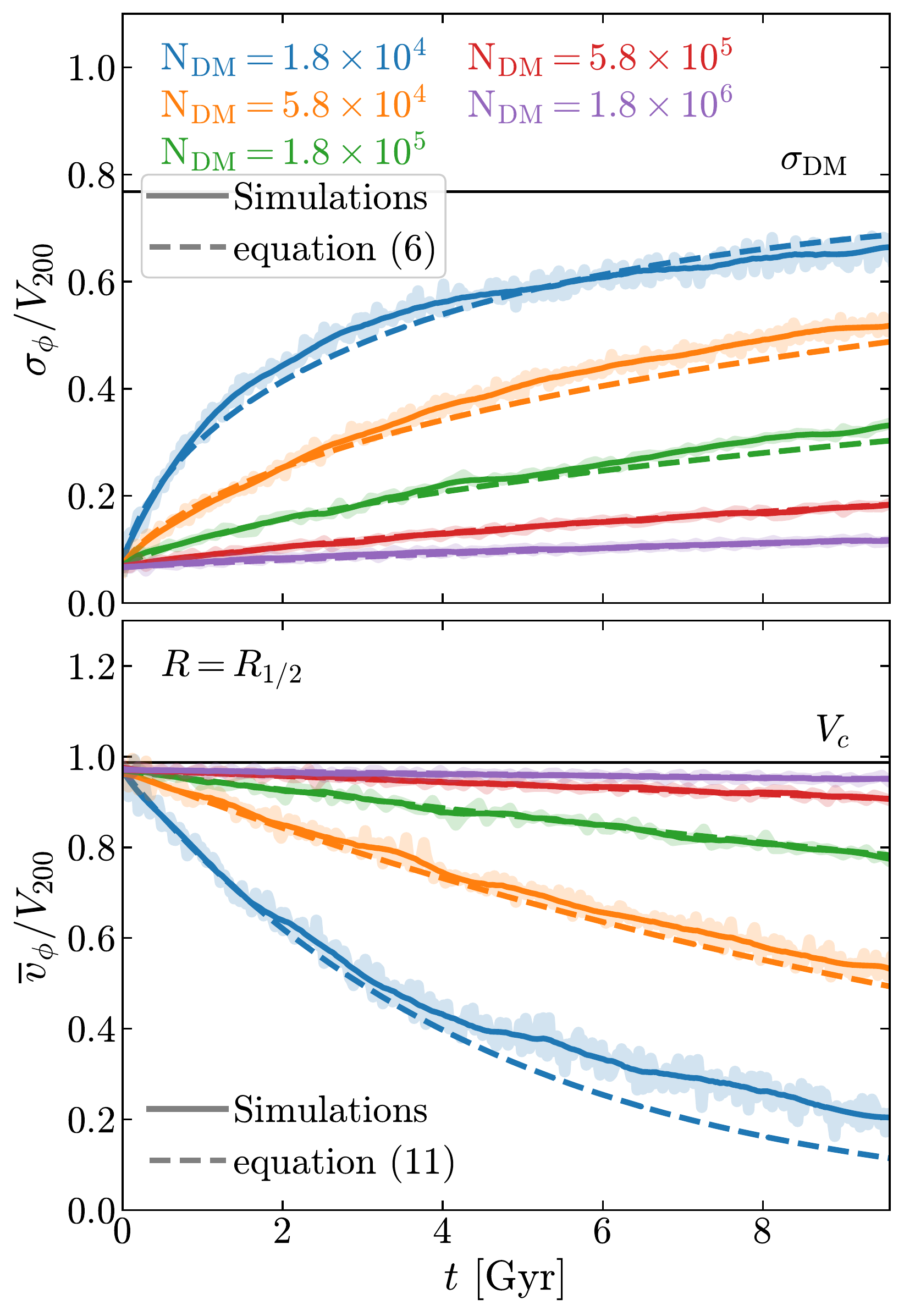}
\caption{Left-hand panels plot radial profiles of the azimuthal velocity dispersion ($\sigma_\phi(R)$; upper panel) and
  the mean azimuthal velocity ($\overline{v}_\phi(R)$; lower panel) after $t=5\,{\rm Gyr}$ for our fiducial galaxy
  models (i.e. $V_{200}=200\,{\rm km/s}$, $\mu=5$, and $c=10$). Profiles are plotted to a minimum radius that encloses 50
  stellar particles. Radii and velocities have been normalised by $r_{200}$ and $V_{200}$, respectively. 
  The dot-dashed black lines in the left-hand panels show the initial velocity dispersion and azimuthal velocity profiles
  for our fiducial model. 
  The solid black line in the upper-left panel corresponds to the one-dimensional velocity dispersion of the Hernquist dark
  matter halo; in the lower panel, the black line shows the {\em total} circular velocity profile of the halo plus the (initial)
  exponential disc. The vertical dotted lines in the left-hand panels indicate the initial value of $R_{1/2}$; circles mark the
  measured values of $R_{1/2}$ at $t=5\,{\rm Gyr}$. The right-hand panels plot the time
  evolution of $\sigma_\phi(R_{1/2})$ (top) and $\overline{v}_\phi(R_{1/2})$ (bottom) measured at the {\em initial} half
  stellar-mass radius of the galaxy.
  Horizontal grey lines in the panels on the right mark the local values of the DM halo velocity dispersion (upper-right panel)
  and total circular velocity (lower-right panel). In all panels, different coloured lines distinguish runs carried out with different
  DM particle masses, as indicated in the legend. The various coloured dashed lines in each panel show the predictions of our empirical
  disc heating model (see \Cref{ssec:model}, \Cref{sec:fitting} and \citealt{Ludlow2021} for details). The dotted lines in the
  lower-left panel show the azimuthal velocity profiles expected from asymmetric drift (i.e. equation~\ref{eq:asymm_drift}), for which the
  stellar particle velocity dispersions were predicted by the disc heating model.}
\label{fig:kinematic-profiles}
\end{figure*}

\citet{Ludlow2021} showed that spurious collisional heating increases the vertical and radial velocity dispersion of stellar disc
particles, as well as the thickness of discs. In Fig.~\ref{fig:kinematic-profiles} we show that the same is true for the
azimuthal velocities of star particles, where in the left-hand panels we plot radial profiles of $\sigma_\phi$ (upper-left) and
$\overline{v}_\phi$ (lower-left) after $t=5\,{\rm Gyr}$, and in the right-hand panels we plot the evolution of $\sigma_\phi$ and
$\overline{v}_\phi$ measured at the {\em initial} value of $R_{1/2}$ (circles in the panels on the left mark the instantaneous values of $R_{1/2}$ at $t=5\,{\rm Gyr}$). Velocities and radii have been normalised by $V_{200}$ and $r_{200}$, respectively,
and results are shown for our fiducial models (i.e. $V_{200}=200\,{\rm km/s}$, $\mu=5$,
and $c=10$) using different coloured lines for different mass resolutions. For comparison, we also plot the
one-dimensional velocity dispersion profile of the DM halo (solid black line, upper-left panel) and the {\em total} circular velocity profile
(solid black line, lower-left panel) of the halo and disc. (Note that neither of these halo 
    profiles evolve noticeably during the course of the simulations.)
The thick, faint lines show the results from our simulations; the thin, dark lines of corresponding colour have been smoothed
for aesthetic purposes.\footnote{In Figs.~\ref{fig:kinematic-profiles} \ref{fig:size-angular-momentum}, \ref{fig:shape-global},
  \ref{fig:predicted-kinematics} and \ref{fig:unpredicted-kinematics}, we use a Savitzky-Golay filter with a width of $2\,{\rm Gyr}$
  to smooth measurements obtained directly from the simulation outputs, and plot these as dark lines.}

These results are reminiscent of those obtained by \citet{Ludlow2021}, and show that the azimuthal velocity dispersion of stellar
particles increases with time as a result of spurious collisional heating. The severity of the effect increases with increasing
DM particle mass (at fixed halo mass) and with decreasing radius, the latter due to the increased density of DM ``perturbers''
closer to the halo centre. Note too that the increase in velocity dispersion is accompanied by a decrease in the mean azimuthal
velocity of stellar particles. This is the well-known phenomenon of increasing ``asymmetric drift'' with increasing velocity dispersion,
as observed for stars in the solar neighbourhood. In the Milky Way, disc stars are gravitationally scattered by past encounters with
satellite galaxies, molecular clouds, spiral arms and other non-axisymmetric features, whereas in our simulations, they are scatted
primarily by individual DM halo particles. In both cases, the scattering randomly perturbs the angular momentum vectors of the disc stars,
thus effectively converting coherent rotation into random motion.

Although the majority of the spurious heating observed in our runs occurs due to collisions between dark matter
    and stellar particles, the self-scattering of stellar particles, particularly in our low-resolution runs, is non-negligible.
    To test the magnitude of the effect, we repeated two of our fiducial runs after replacing the ``live'' Hernquist
    DM halo with a static potential. For our lowest-resolution run ($m_{\rm DM}=10^8\,{\rm M_\odot}$; $N_\star=900$), we find
    that the dispersion in the azimuthal velocities of disc particles (measured at $R=R_{1/2}$) is $\sigma_\phi\approx 31\,{\rm km/s}$
    after $t=5\,{\rm Gyr}$; this is a factor of $\approx 2.8$ larger than the initial velocity dispersion, but a factor of $\approx 3.7$
    lower than in the corresponding run carried out with a live halo. For our second-lowest run ($m_{\rm DM}=10^{7.5}\,{\rm M_\odot}$;
    $N_\star=2900$), we find $\sigma_\phi=23\,{\rm km/s}$ (again measured at $R=R_{1/2}$ after $t=5\,{\rm Gyr}$), a factor of
    $\approx 3.3$ lower than the live halo case.

Clearly the spurious heating of stellar motions by dark matter halo particles has implications for the evolution of a number of
other dynamical properties of galactic
discs, such as their sizes, rotation velocities, angular momenta, and shapes. We discuss each of these in the sections that follow. 
First however, we note that the dynamical effects of collisional heating can be described reasonably well by a simple empirical model,
shown as dashed coloured lines in each panel of Fig.~\ref{fig:kinematic-profiles}. This model is a slightly modified and extended
version of the one first presented by \citet{Ludlow2021}; it is described below in \Cref{ssec:model} and in more detail in \Cref{sec:fitting}.

\subsection{Galaxy scaling relations}
\label{ssec:scaling-relations}

\begin{figure*}
  \includegraphics[width=1\textwidth]{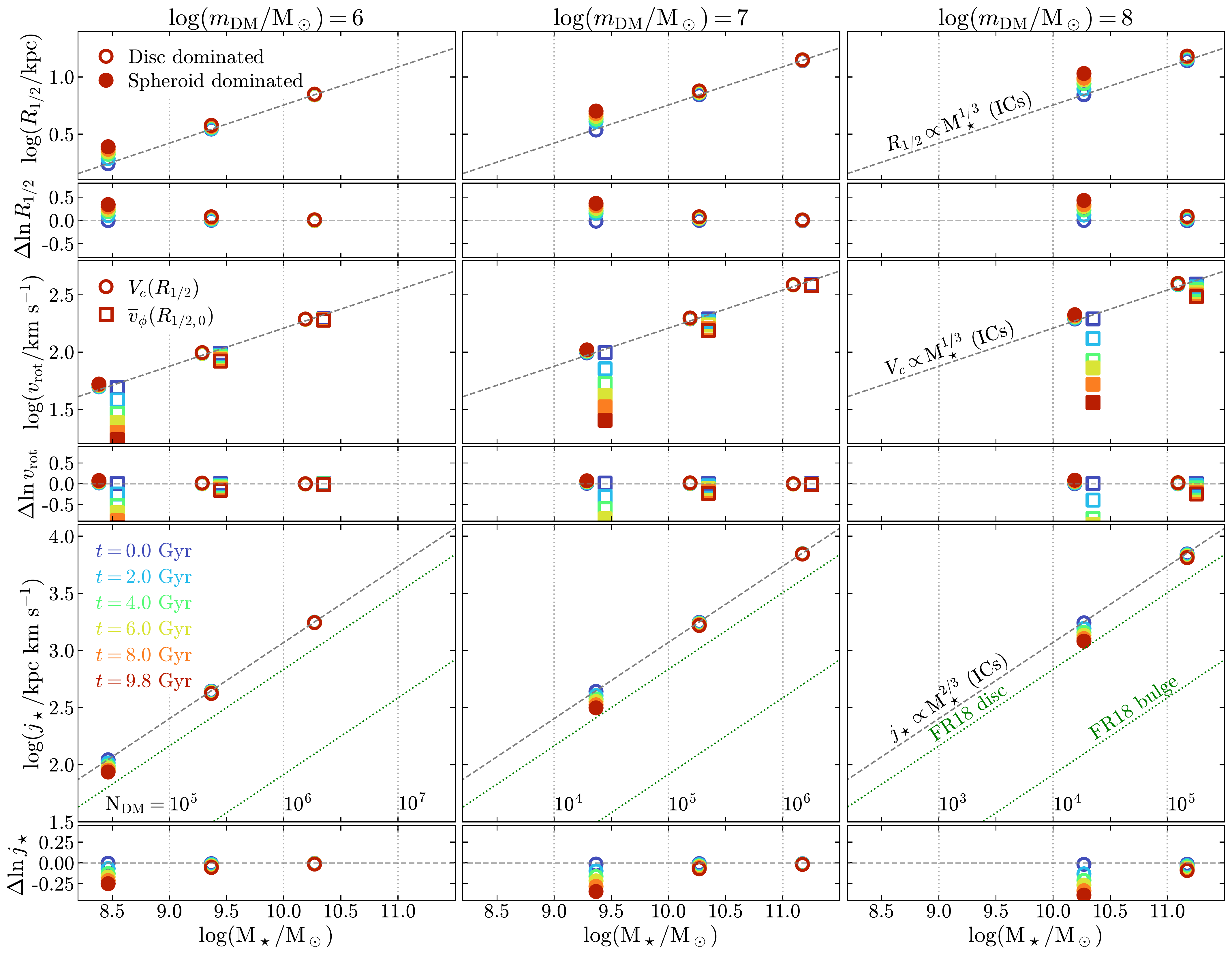}
  \caption{The impact of collisional heating on the relation between $R_{1/2}$ (the half mass radius; top panels),
    $v_{\rm rot}(R_{1/2})$ (the rotational velocity at $R_{1/2}$; middle panels), $j_\star=J_\star/M_\star$
    (i.e. the total specific stellar angular momentum; bottom panels) and stellar mass. We use two independent measures of $v_{\rm rot}$:
    one is the total circular velocity at $R_{1/2}$ (circles); the other is the mean azimuthal velocity of
    stellar particles at the initial value of $R_{1/2}$ (squares). Each column shows all three relations obtained
    from simulations run with a particular dark matter particle mass, $m_{\rm DM}$, which increases from left to right. The
    various sets of points in each panel correspond to haloes of different $V_{200}$ (vertical dotted lines
    indicate the values of ${\rm M_\star}$ corresponding to a few characteristic $N_{\rm DM}$, assuming $\mu=5$ and
    $f_\star=0.01$) and their colour indicates the output time, as indicated in the legend in the lower-left panel.
    For well-resolved haloes ($N_{\rm DM}\gtrsim 10^6$), the evolution of $R_{1/2}$, $v_{\rm rot}$, and $j_\star$ is
    insignificant and the points largely overlap; for poorly-resolved haloes ($N_{\rm DM}\lesssim 10^5$), however, evolution can be
    significant. For example, we use filled and open symbols to distinguish disc-dominated systems from spheroids, respectively,
    where the latter are defined as galaxies for which twice the counter-rotating mass fraction exceeds 0.5. The grey dashed
    lines in each panel show the expected scaling relations given by our initial conditions; the residuals in the lower panels
    are calculated with respect to these lines. For reference, the bottom row of panels include the parallel relations
    obtained by \citet{Fall2018} for observed discs and bulges (green dotted lines).}
  \label{fig:scaling-relations}
\end{figure*}

The results above imply that collisional heating drives spurious evolution in the structure and dynamics of
disc galaxies. We investigate this in Fig.~\ref{fig:scaling-relations}, where we plot a subset of our model galaxies on several standard
scaling laws. Different columns separate runs carried out using different DM particle masses:
from left to right, $m_{\rm DM}=10^6\,{\rm M_\odot}$, $m_{\rm DM}=10^7\,{\rm M_\odot}$, $m_{\rm DM}=10^8\,{\rm M_\odot}$,
respectively. The different sets of points in each panel distinguish the different $V_{200}$ values
(at fixed $m_{\rm DM}$, these correspond to different $N_{\rm DM}$; we indicate using vertical
dotted lines the values of ${\rm M_\star}$ corresponding to a few characteristic values of $N_{\rm DM}$, assuming
$\mu=5$ and $f_\star=0.01$).
We use points of different colour for different output times (as labelled in the lower-left panel).

From top to bottom, different rows show the size-mass relation, the relation between
rotational velocity and stellar mass (i.e., the Tully-Fisher \citeyear{Tully1977} relation),
and the specific angular momentum-stellar mass relation
(i.e., the \citealt{Fall1983} relation), respectively.
The various grey dashed lines in each panel show the corresponding relations
obtained from the initial conditions of our simulations: from top to bottom, $R_{1/2}\varpropto {\rm M}_\star^{1/3}$,
$v_{\rm rot}\varpropto {\rm M}_\star^{1/3}$, and $j_\star\varpropto {\rm M}_\star^{2/3}$, respectively. Each panel is accompanied
beneath by a smaller ``residual'' panel, which shows the measured departure of our simulations from these initial relations.

The upper panels of Fig.~\ref{fig:scaling-relations} demonstrate that poorly-resolved discs experience a significant
increase in size due to spurious heating, a result previously discussed by \citet[][see also \citealt{Revaz2018, Ludlow2020}]{Ludlow2019}.
The effect is most severe for models with $N_{\rm DM}\lesssim {\rm a\, few}\times 10^5$: in our lowest-resolution
runs for example, corresponding to $N_{\rm DM}\approx 2\times 10^4$, $R_{1/2}$ increased by about 60 per cent after $9.8 \,{\rm Gyr}$;
for $N_{\rm DM}\approx 5 \times  10^4$ (not shown in Fig.~\ref{fig:scaling-relations}), $R_{1/2}$ increased by only about 10 per cent over the
same time interval. For $N_{\rm DM} \gtrsim 10^6$, however, the half-mass size evolution is negligible,
exhibiting less than 2 per cent growth over the course of the simulation. Similar results are obtained for $R_{1/4}$ and $R_{3/4}$, enclosing
one quarter and three quarters of the galaxy's stellar mass, respectively.

The middle panels of Fig.~\ref{fig:scaling-relations} plot the relation between rotational velocity, $v_{\rm rot}$,
and stellar mass. The two sets of points in each panel correspond to two complimentary measurements of
$v_{\rm rot}$: in one case (circles) $v_{\rm rot}=V_c(R_{1/2})$ is taken to be the total circular velocity at $R_{1/2}$; in
the other case (squares), $v_{\rm rot}=\overline{v}_\phi(R_{1/2, 0})$ is the mean azimuthal velocity of stellar particles
measured within a cylindrical aperture centred on the {\em initial} value of $R_{1/2}$. We note that many observational
studies define galaxy rotation velocities by averaging the outer-most points of their rotation curves, typically between one and
${\rm a \, few}\times R_{1/2}$, which minimises the scatter in the Tully-Fisher relation compared to other velocity measurements
\citep[e.g.][]{Verheijen2001, Lelli2019}. Although this differs somewhat from our estimates of $v_{\rm rot}$, we note that good
agreement between the Tully-Fisher relations of observed and simulated galaxies can be obtained using $v_{\rm rot}=V_c(R_{1/2})$ for the latter
\citep[e.g.][]{Ferrero2017}.

Clearly, $v_{\rm rot}$ is affected by spurious collisional heating, but the magnitude of the effect depends strongly on
how it is calculated: it is much less severe when estimated from the total circular velocity profile than from the mean
azimuthal velocity of stellar particles. This is because
our galaxy discs have relatively low mass fractions ($f_\star = 0.01$) and therefore do not contribute much to $V_{\rm c}$, but
also due to the fact that $R_{1/2}$ probes a radial range over which $V_{\rm DM}$ rises relatively slowly; the small increase
in $v_{\rm rot}= V_{\rm c}(R_{1/2})$ is therefore primarily due to the spurious growth of $R_{1/2}$ probing slightly higher DM
circular velocities.

If $v_{\rm rot}$ is instead estimated using the mean azimuthal velocities of star particles, then its evolution can be significant.
Models with $N_{\rm DM}={\rm a\, few}\,\times 10^4$, for example, see $v_{\rm rot}$ reduced by about a factor of 4 relative to
its initial value over the course of the simulations. Even models with $N_{\rm DM}={\rm a\, few\,\times 10^5}$ are not immune:
for those, $v_{\rm rot}$ drops by about 26 per cent over ${\rm 9.8\,Gyr}$. This is in fact a natural consequence of the collisional
heating of disc galaxies, which harbour a large fraction of their initial kinetic energy
in ordered rotational motion. In addition to {\em heating} stellar particle orbits -- which occurs primarily as a result of
energy equipartition and mass segregation -- collisions between stellar and DM particles 
perturb their motions, thereby converting ordered rotation into velocity dispersion.

It is therefore not surprising that simulated stellar discs also tend to lose angular momentum as a result of spurious
    collisional heating (as discussed previously by \citealt{Governato2004}).
The effect, shown in the bottom panels of Fig.~\ref{fig:scaling-relations}, is however most problematic for runs
with $N_{\rm DM}\lesssim {\rm a\, few}\times 10^5$, for which up to 30 per cent of the disc's initial angular momentum can be
lost over $9.8\,{\rm Gyr}$. We note, however, that
in none of our simulations is the loss of angular momentum sufficient to migrate the points from the relation occupied by ``discs''
to the one occupied by ``bulges'' in both observations and cosmological simulations (the diagonal dotted lines in the lower
panels show the relations for observed discs and bulges obtained by \citealt{Fall2018}).

Although this effect may seem reminiscent of the angular momentum catastrophe, which plagued early simulations of
    galaxy formation, it is not the same. The angular momentum catastrophe has its origin in the efficient cooling of
    baryons in high-redshift protohaloes, which
    subsequently merge and lose angular momentum due to dynamical friction, resulting in present-day galaxies with too
    little angular momentum and unrealistically small sizes \citep[see, e.g.,][]{NavarroSteinmetz2000}.
    Although our simulated discs lose angular momentum to their surrounding halo as a result of collisional heating,
    they do not lose energy; in fact, they {\em expand} as result of the approach toward energy equipartition.

Fig.~\ref{fig:scaling-relations} highlights one consequence of collisional heating, already apparent in
Figs.~\ref{fig:projections-low}, ~\ref{fig:projections-med}, and \ref{fig:projections-hi}: it
drives morphological changes in disc galaxies by converting thin, rotationally supported discs into dispersion-dominated
spheroids. This is emphasised in the various panels of Fig.~\ref{fig:scaling-relations}, where we plot disc-dominated systems using filled symbols
and dispersion-dominated ones using open symbols (note that the latter are defined as galaxies for which twice the
counter-rotating mass fraction is $S/T \geq 0.5$; otherwise they are disc dominated). 

\begin{figure}
  \includegraphics[width=0.5\textwidth]{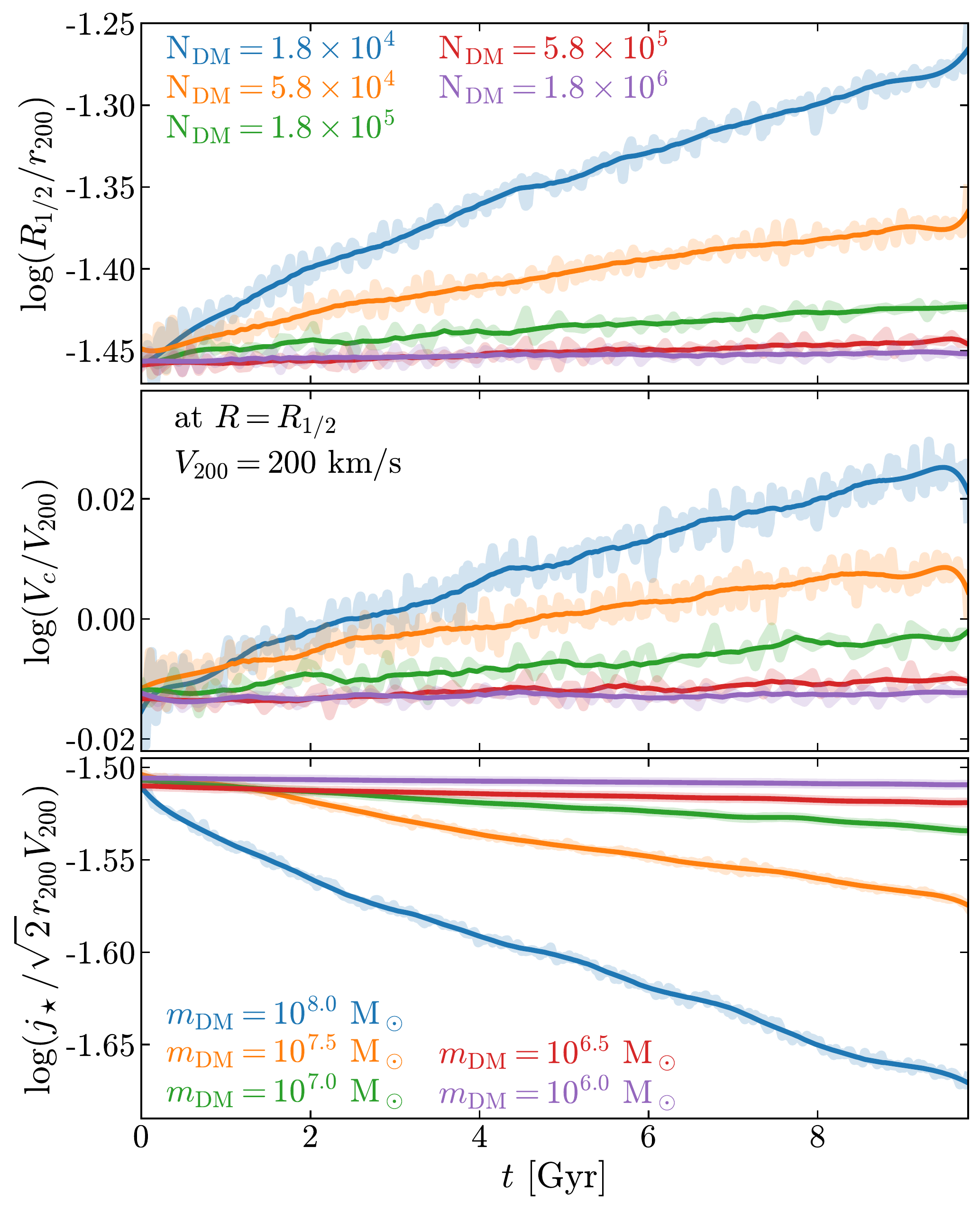}
  \caption{From top to bottom, respectively, we plot the evolution of the half-stellar mass radius, $R_{1/2}$,
    the total circular velocity at the {\em initial} value of $R_{1/2}$ ($V_c$ at the instantaneous value of $R_{1/2}$ is
    often used as a proxy for the rotational velocity of simulated galaxies),
    and total specific stellar angular momentum, $j_\star$. The line colours represent runs carried out with different dark matter
    particle masses, as indicated in the legend. The thick, light-coloured lines show the results from our simulations; the thin,
    dark lines show the results after smoothing the simulation outputs with a Savitzky-Golay filter for presentation purposes. All
    results are shown for our fiducial runs with ${\rm V_{200}}=200\,{\rm km/s}$ and $\mu=5$.}
  \label{fig:size-angular-momentum}
\end{figure}

In Fig.~\ref{fig:size-angular-momentum} we plot the time dependence of each of these quantities obtained for our fiducial
galaxy models (i.e. $V_{200}=200\,{\rm km/s}$ and $\mu=5$). The upper panel plots the evolution of the stellar
half-mass radius (normalised by $r_{200}$); the middle panel plots the total circular velocity at $R_{1/2}$ (normalised
by $V_{200}$; we considered the evolution of $\overline{v}_\phi$ separately in Fig.~\ref{fig:kinematic-profiles});
and the bottom panel plots the evolution of the specific angular momentum of all stellar particles (normalised
by $\sqrt{2}\,V_{200}\,r_{200}$). 

Although the impact of collisional heating on galaxy scaling relations is often small (at least for galaxies
occupying well-resolved DM haloes), it is systematic, and becomes increasingly problematic 
with time. It also disproportionately affects poorly-resolved systems, for which the effect
clearly cannot be ignored. For cosmological simulations that adopt a uniform mass resolution for DM particles, there
will always be a halo mass scale below which these effects are potentially important. We will address this in a companion
paper, but for the remainder of this paper we focus our attention on how spurious collisional heating alters the kinematics and
morphologies of isolated, idealised disc galaxies. 

\subsection{The spurious evolution of disc galaxy morphology}
\label{ssec:measurements}

\begin{figure}
  \includegraphics[width=0.5\textwidth]{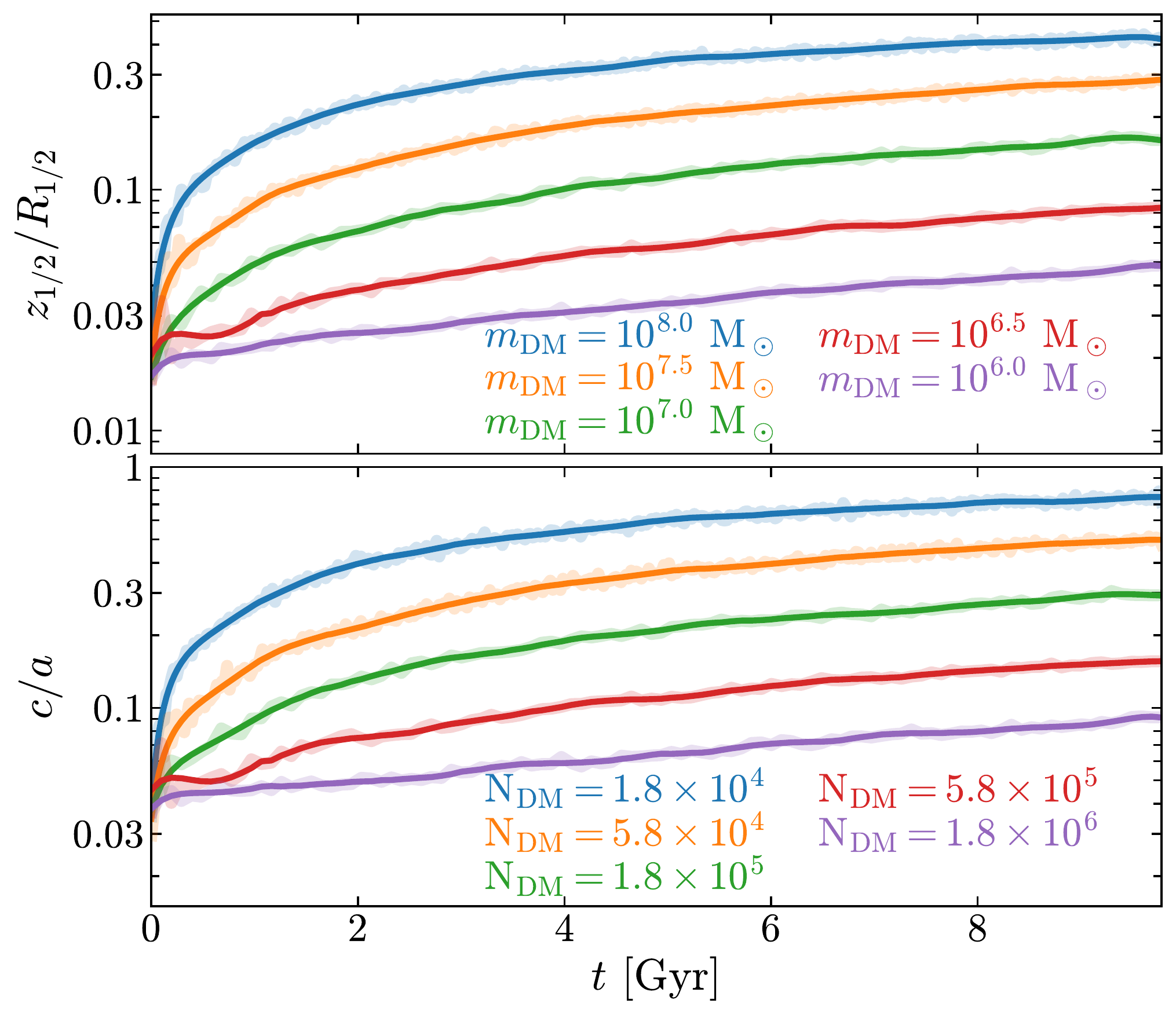}
  \caption{The time evolution of the aspect ratio of our fiducial galactic discs (i.e. $V_{200}=200\,{\rm km/s}$ and $\mu=5$).
    The upper panel shows the ratio of the half-mass scale height to the half-mass scale length, i.e. $z_{1/2}/R_{1/2}$;
    in the lower panel we plot the time-dependence of the minor-to-major
    axis ratio, $c/a$, estimated from the spatial distribution of all stellar particles. Lines of different colour correspond to
    runs of different resolution, as indicated. These measurements confirm and quantify the morphological evolution of discs already
    evident from their visual appearance (See Figs.~\ref{fig:projections-low} to \ref{fig:projections-hi}).}
  \label{fig:shape-global}
\end{figure}

As alluded to above (Fig.~\ref{fig:scaling-relations}), spurious collisional heating can alter the morphologies of simulated disc galaxies,
transforming them into spheroids. 
In this section, we quantify these effects using kinematic and structural estimates of galaxy morphology. 
For simplicity, we only present results for our fiducial runs (i.e. $V_{200}=200\, {\rm km/s}$, $\mu=5$),
but have verified that our conclusions apply to all of our simulated discs.
	
\subsubsection{The evolving shapes of galactic discs}
\label{ssec:shapes}

Fig.~\ref{fig:shape-global} plots the time evolution of two complimentary structural estimates of galaxy shape. In the
upper panel, we plot the aspect ratio of the half-mass height to the half-mass length, i.e. $z_{1/2}/R_{1/2}$.
As with previous plots, runs carried out with different $m_{\rm DM}$ are plotted using different coloured lines.

Initially, all models have $z_{1/2}/R_{1/2}\approx 0.016$, as expected for thin discs. 
However, all discs experience an increase in $z_{1/2}$ relative to $R_{1/2}$, with the rate of increase proportional to $m_{\rm DM}$. 
For our lowest-resolution runs the aspect ratio initially increases very rapidly, but later slows as the vertical velocity dispersion
of stellar particles approaches the maximum value dictated by the local one-dimensional velocity dispersion of the DM halo (the latter imposes
a maximum vertical scale height through the hydrostatic equilibrium equation; see \citealt{Ludlow2021} for details). 
This suggests that the vertical structure of discs is much more vulnerable to collisional heating than their radial structure, or surface density profile.
To quote a few numbers, our lowest-resolution model (blue lines) has $z_{1/2}/R_{1/2} = 0.16, 0.24, 0.32, 0.43$ after $t=1, 2, 4$ and $9.8\,{\rm Gyr}$
respectively; i.e. the final value of $z_{1/2}/R_{1/2}$ is roughly a factor of 27 larger than its initial value.
Increasing the DM mass resolution reduces the effect, but does not eliminate it. For example, for $m_{\rm DM}=10^7\,{\rm M_\odot}$ ($N_{\rm DM} = 1.8 \times 10^5$),
we find $z_{1/2}/R_{1/2}\approx 0.2$ by the end of the simulation, about an order of magnitude larger than at $t=0$. Even for our highest-resolution galaxy
($N_{\rm DM} = 1.8 \times 10^6$) the disc aspect ratio grows noticeably with time, increasing by a factor of $\approx 3$ over 9.8 Gyr, although its disc-like
properties are preserved overall.

The lower panels plot the minor-to-major axis ratios, $c/a$, for the same set of models. For our highest-resolution run, which retains a disc-like appearance
after $t=9.8\,{\rm Gyr}$ (as easily seen in the left-most column of Fig.~\ref{fig:projections-med}), $c/a$ increases by a factor of roughly 2.5 over the same
time interval, resulting in a final axis ratio of $c/a\approx 0.09$. This is a considerable change despite the galaxy's halo being resolved with
$N_{\rm DM}\gtrsim 10^6$ particles. As the mass resolution decreases the discs become substantially more spheroidal. For example, for
$m_{\rm DM}=10^{7}\,{\rm M_\odot}$ ($N_{\rm DM}=1.8\times 10^5$), the final axis ratio is $c/a\approx 0.3$, and for our lowest-resolution run
($N_{\rm DM}=1.8\times 10^4$), $c/a\approx 0.7$. This galaxy resembles a spheroid more than a disc.

\subsubsection{Stellar kinematic indicators of galaxy morphology}
\label{ssec:kinematic-indicators}

\begin{figure}
  \includegraphics[width=0.5\textwidth]{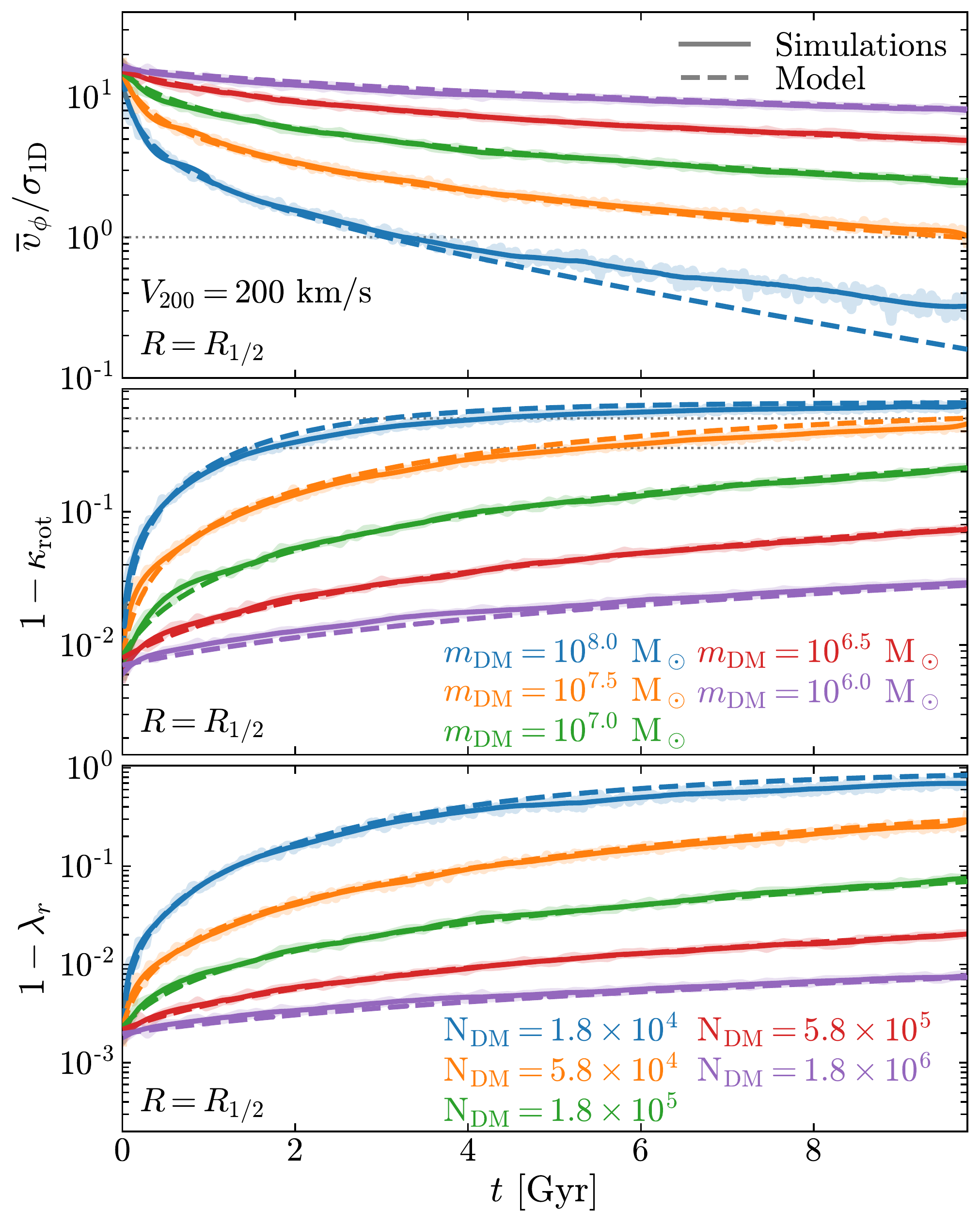}
  \caption{The evolution of three kinematic indicators of galaxy morphology, including $\overline{v}_\phi/\sigma_{\rm 1D}$
    (i.e. the ratio of rotation to dispersion velocities; upper panel), $\kappa_{\textrm{rot}}$ (the fraction of kinetic energy in
    rotational motion; middle panel), and $\lambda_r$ (the stellar spin parameter; lower panel). As in previous figures, line colours 
    differentiate runs carried out with different DM particle masses. Note that each of these quantities was measured in a cylindrical
    shell of logarithmic width $\Delta\log R=0.2$ centred on the {\em initial} half stellar-mass radius of the disc, $R_{1/2}$. The
    dashed coloured lines show the predictions of our empirical disc heating model (See \Cref{ssec:model}, \Cref{sec:fitting} for a
    detailed discussion of our collisional heating model).
    The horizontal dotted line in the upper panel at $\overline{v}_\phi / \sigma_{\rm 1D} = 1$ delineates rotation- and dispersion-dominated
    systems. In the middle panel dotted lines at $\kappa_{\rm rot} = 0.7$ and 0.5 distinguish disc dominated ($\kappa_{\rm rot}\geq 0.7$)  and
    spheroid-dominated ($\kappa_{\rm rot}\leq 0.5$) systems, respectively.}
  \label{fig:predicted-kinematics}
\end{figure}

In Fig.~\ref{fig:predicted-kinematics} we plot the evolution of several kinematics-based indicators of galaxy morphology, again
limiting our results to our fiducial runs. From top to bottom, different panels correspond to $\overline{v}_\phi/\sigma_{\rm 1D}$
(i.e. the ratio of rotation-to-dispersion velocities), $\kappa_{\rm rot}$ (i.e. equation~\ref{eq:kappa-data}, which measures the
fraction of stellar kinetic energy in rotation), and $\lambda_r$ (equation~\ref{eq:lambda-data}, the disc spin parameter). In order
to more easily compare these results to a simple analytic model (described later in \Cref{ssec:model}), all quantities were measured
in a cylindrical aperture of width $\Delta\log R=0.2$ centred on the initial value of $R_{1/2}$, although other radii $R_f$ could
also have been used. As in the previous two figures, different coloured lines distinguish runs with different mass resolution.
Note that qualitatively similar results are obtained using all stellar particles rather than those occupying a particular radial shell. 

In line with previous results, the morphologies of our disc galaxies -- as quantified by these kinematic
quantities -- move progressively from disc-like to spheroid-like with time. And as before, the rate of morphological transformation
is strongly correlated with mass resolution. All of our models have a relatively high initial ratio of rotation to dispersion velocities, i.e.
$\overline{v}_\phi/\sigma_{\rm 1D}\approx 17$, but collisional
heating converts ordered rotational motion to velocity dispersion causing $\overline{v}_\phi/\sigma_{\rm 1D}$ to decrease steadily with
time, an effect that is weak but noticeable in our highest-resolution runs and substantial in our lowest-resolution runs.
For example, in order of decreasing mass resolution and after approximately 8 Gyr of evolution,
our simulated discs have $\overline{v}_\phi/\sigma_{\rm 1D}\approx 8.8$, 5.4, 2.8, 1.3 and 0.4, respectively.
Our poorest-resolved galaxy becomes dispersion dominated (i.e. has $\overline{v}_\phi/\sigma_{\rm 1D}\lesssim 1$) after only $\approx 3.3\,{\rm Gyr}$.
Even in our intermediate-resolution model (green lines in Fig.~\ref{fig:predicted-kinematics}),
in which the galaxy's DM halo is resolved with roughly $2\times 10^5$ DM particles, $\overline{v}_\phi/\sigma_{\rm 1D}$ drops from $\approx 17$
at $t=0$ to $\approx 2.5$ after $t=9.8\,{\rm Gyr}$. This suggests that resolving thin, rotationally supported stellar discs requires their
DM haloes to be resolved with at least $10^6$ particles, an estimate that we verify quantitatively in \Cref{ssec:cosmosims}.

The results plotted in the middle and lower panels, for $\kappa_{\rm rot}$ and $\lambda_r$, respectively, tell a similar story: Collisional
heating drives galactic discs from kinematically cold structures, dominated by ordered rotational motion, to warm
(or in extreme cases {\em hot}) systems supported substantially by the random motions of their stellar particles. 
Our lowest-resolution model, for example, falls below the $\kappa_{\rm rot}\gtrsim 0.5$ \citep[e.g.][]{Sales2012} threshold often used
to classify discs after only $t=4.1\,{\rm Gyr}$; and our second-lowest-resolution model is intermediate between disc- and bulge-dominated
(i.e. $0.7 > \kappa_{\rm rot} > 0.5$), with $\kappa_{\rm rot}=0.52$ after $t=9.8\,{\rm Gyr}$.

\begin{figure}
  \includegraphics[width=0.48\textwidth]{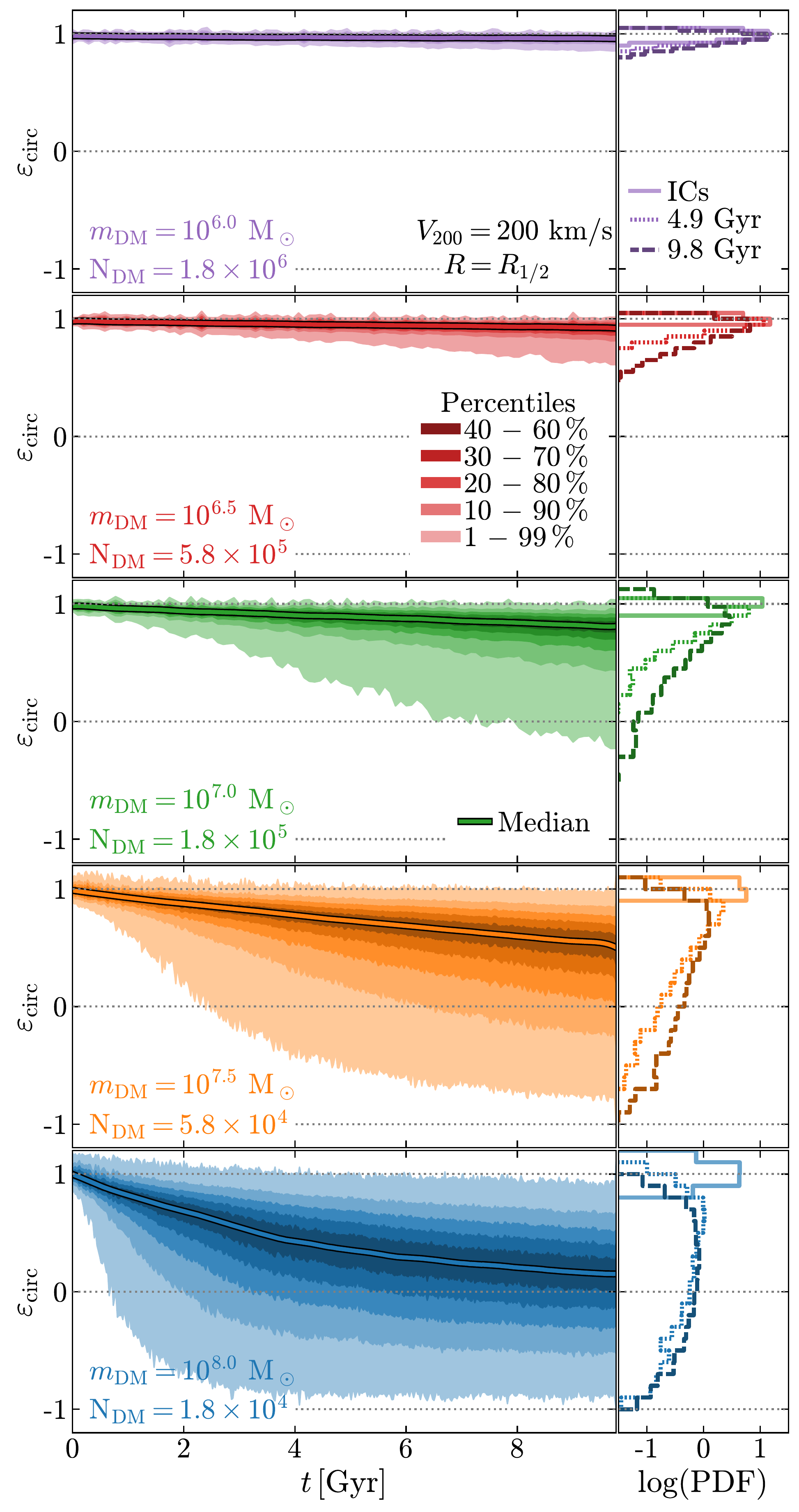}
  \caption{Evolution of the circularity parameter, $\varepsilon_{\rm circ}$, for our fiducial runs (i.e. $V_{200}=200\,{\rm km/s}$,
    $\mu=5$). From top to bottom, different panels correspond to increasing DM particle mass and decreasing particle number; the 
    values of each are indicated in the appropriate panel.
    The solid coloured lines correspond to the median values of $\varepsilon_{\rm circ}$ obtained for all stellar particles. From lightest to
    darkest, the various shaded regions enclose the $1-99$, $10-90$, $10-80$, $30-70$ and $40-60$ inter-percentile scatter.
    Smaller panels to the right show the normalised probability distributions of $\varepsilon_{\rm circ}$ at $t=0$, and after $t=4.9$
    and 9.8 Gyr of evolution.}
  \label{fig:circularity}
\end{figure}

Another quantity often used to diagnose galaxy morphology is the orbital circularity parameter, $\varepsilon_{\rm circ}\equiv j_z/j_{\rm c}(E)$.
This quantity can be calculated for each
stellar particle, and can therefore be used to decompose a simulated galaxy into distinct components: e.g., disc stars are commonly
defined by the subset of particles with $\varepsilon_{\rm circ}\gtrsim 0.7$ \citep[e.g.][]{Aumer2013, Grand2017, Joshi2020}, and the mass of
a galaxy's spheroidal component as twice the mass fraction having $\varepsilon_{\rm circ}<0$ \citep[e.g.][]{Abadi2003}.
Although more sophisticated methods to dynamically decompose galaxies exist \cite[see e.g.][]{Scannapieco2009, Domenech-moral2012, Obreja2016},
simple criteria based on circularity thresholds such as those described above are also commonplace, so we choose this straightforward method
to assign disc-to-total ($D/T$) and spheroid-to-total ($S/T$) mass ratios to our simulated discs. 

In Fig.~\ref{fig:circularity} we plot the evolution of the median circularity parameter of stellar particles, $\varepsilon_{\rm circ} = j_z / j_c(E)$,
in our fiducial runs (thick lines), i.e $V_{200}=200\, {\rm km/s}$, $\mu=5$. Simulations adopting different particle masses are shown in separate panels, but use
the same colour-coding as previous plots. Shaded regions highlight the inter-percentile ranges corresponding to $1-99$
(lightest), $10-90$, $20-80$, $30-70$ and $40-60$ (darkest). Smaller panels to the right of the main ones show the distributions
of $\varepsilon_{\rm circ}$ at $t=0$ and after $t=4.9$ and 9.8 Gyr (darker to lighter lines). Note that at $t=0$, virtually all stellar
particles are co-orbiting, as expected for pure disc galaxies. As time goes on, however, spurious collisional heating disturbs an increasing
number of stellar particle orbits, resulting in a slow but systematic decrease in the median value of $\varepsilon_{\rm circ}$ and a corresponding
increase in its dispersion. After several Gyr of evolution, a substantial fraction of the stellar mass of our poorly-resolved haloes
has been scattered onto counter-rotating orbits, most likely originating from repeated large-angle deflections during encounters with DM particles. 
Indeed, a small minority (about 3 per cent) of stellar particles in our lowest-resolution run find themselves on counter-rotating
orbits {\em in the disc plane} (i.e. they have $\varepsilon_{\rm circ} < -0.7$).

\begin{figure}
  \includegraphics[width=0.45\textwidth]{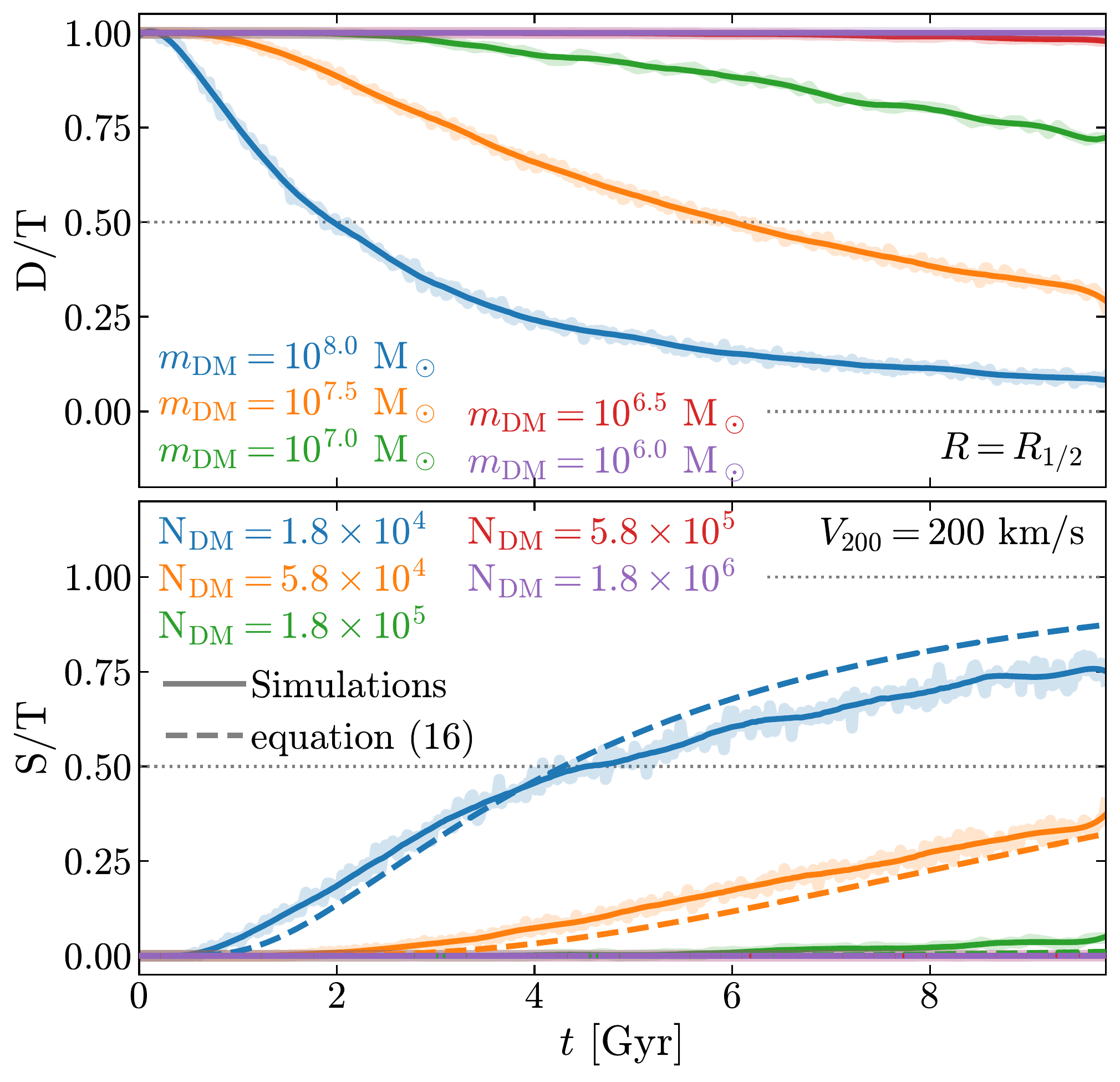}
  \caption{Estimates of the disc-to-total ($D/T$; upper panel) and spheroid-to-total ($S/T$; lower panel) mass ratios for our fiducial stellar discs
    (i.e. ${\rm V_{200}=200\, km/s}$ and $c=10$). As in previous figures, different colours correspond to runs carried out with different $m_{\rm DM}$,
    as indicated. Note that the ratios $D/T$ and $S/T$ are estimated from the fraction of stellar particles with $\varepsilon_{\rm circ} >0.7$ and twice
    the fraction of stellar particles with $\varepsilon_{\rm circ} < 0$, respectively (or equivalently, twice the mass fraction with $v_\phi < 0$). The
    dashed lines in the lower panel show the evolution of $S/T$ as predicted by our disc heating model (see Section~\ref{ssec:model}; equation~\ref{eq:ST2}).}
  \label{fig:unpredicted-kinematics}
\end{figure}

The galaxies in our two lowest-resolution simulations
(bottom-most two panels) eventually develop median circularity parameters that dip
below the threshold typically associated with disc stars, i.e. $\varepsilon_{\rm circ} \approx 0.7$. Using traditional diagnostics, a large fraction of their
mass (95 and 76 per cent for the lowest- and second-lowest-resolutions models, respectively)
would therefore not be associated with a disc; it would
instead be associated with a bulge/spheroid or, potentially, a thick disc. We emphasise this in Fig.~\ref{fig:unpredicted-kinematics},
where we plot the time dependence of the disc-to-total ($D/T$; upper panel; defined as the mass-fraction with $\varepsilon_{\rm circ}\geq 0.7$) and
spheroid-to-total ($S/T$; lower panel; defined as two times the mass fraction with $\varepsilon_{\rm circ}\leq 0$) ratios for each model. Only the
two highest-resolution runs retain disc and spheroid fractions of $D/T \approx 1$ and $S/T \approx 0$; our lowest-resolution
run is spheroid-dominated after $\approx 4.5\,{\rm Gyr}$.

\section{An empirical model for spurious collisional heating and its implications for galactic disc morphology}
\label{sec:model}

Many of the numerical results presented in \Cref{ssec:measurements} can be reproduced by a model that describes the
evolution of the velocity dispersion and average azimuthal velocity of stellar disc particles. As we show below, this
can be achieved using an existing analytic description of gravitational scattering \citep[][]{Lacey1985, Ludlow2021}, provided it is
suitably modified to overcome limitations arising from its simplifying assumptions. In Section~\ref{ssec:model} we describe how this can be
achieved, and in Section~\ref{ssec:cosmosims} we discuss the implications of spurious collisional heating for modelling
disc galaxy morphology in cosmological simulations. 

\subsection{An empirical model for spurious collisional heating}
\label{ssec:model}

We follow \citet{Ludlow2021} and use the collisional disc heating rates derived analytically by \citet[][]{Lacey1985} as a
starting point for our empirical model. We present below the equations required to implement the model, and refer the interested
reader to those papers and our Appendix \ref{sec:fitting} for a more detailed discussion.

In the limit $\sigma_i(R)\ll\sigma_{\rm DM}(R)$, the local heating rate of stellar particles results in a incremental
increase in their velocity dispersion that may be approximated by (see \citealt{Ludlow2021})
\begin{equation}
  \frac{\Delta \sigma_i^2}{\Delta t}=\sqrt{2}\,\pi\,\ln\Lambda\,\frac{G^2\,\rho_{\rm DM}\,m_{\rm DM}}{V_{200}}\times k_i\,\biggr(\frac{\rho_{\rm DM}}{\rho_{200}}\biggl)^{\alpha_i}.
  \label{eq:lin}
\end{equation}
Here $\ln\Lambda$ is the Coulomb logarithm, $k_i$ is a dimensionless free parameter that determines the normalisation
of the heating rate, and $\alpha_i$ is a free parameter that governs its dependence on $\rho_{\rm DM}$, 
the local density of the DM halo. The ``$i$'' subscripts on $k_i$ and $\alpha_i$ indicate that their values
differ for the different velocity components, but they do not depend on any intrinsic properties of the halo, or on
radius or time. For $\alpha_i=0$, equation~(\ref{eq:lin}) reproduces the linear density dependence of the collisional heating rate
obtained analytically by \citet[][]{Lacey1985}, with one important difference: it implies a local heating rate that depends inversely
on $V_{200}$, which differs from the $\sigma_{\rm DM}^{-1}$ dependence obtained by \citet[][]{Lacey1985} and adopted in the
\citet{Ludlow2021} model.\footnote{The velocity dependence implied by equation~(\ref{eq:lin}) was determined empirically by
  fitting the model not only to our fiducial runs, but also to those that vary the halo's concentration. By including the latter,
  our models were sensitive to a broader range of halo velocity dispersions than would be the case if fitted to our fiducial runs
  alone, which was the method employed by \citet{Ludlow2021}. For that reason, we provide in \Cref{table:best-fit-values} updated
  model parameters for $\sigma_R$ and $\sigma_z$.}

Although the value of $k_i$ can in principle be calculated explicitly once a velocity distribution for the DM is
specified, we prefer to treat it as a free parameter. This allows us to avoid uncertainties that may
arise due to inaccuracies of the assumed velocity distribution, and offers some freedom when assigning a
value to the Coulomb logarithm, which itself depends weakly on the detailed properties of the disc and halo.
In practice, we treat the combined term $k_i\, \ln\Lambda$ as a free parameter when obtaining our best-fit models.

As discussed by \citet{Ludlow2021}, equation~(\ref{eq:lin}) breaks down when $\sigma_i\approx \sigma_{\rm DM}$, which occurs at late times
for our lowest-resolution runs. Indeed, energy equipartition driven by collisions between stellar and DM particles results
in maximum asymptotic stellar velocity dispersion of $\sigma_{i,{\rm max}}\approx \sqrt{\mu}\times\sigma_{\rm DM}$. Strictly speaking, this can
only be achieved for $\mu\lesssim 2$ since, according to the virial theorem, $\sigma_{i,{\rm max}}=\sqrt{2}\times \sigma_{\rm DM}$.
\citet{Ludlow2021} verified these analytic expectations, but also showed that the typical timescale over which they are reached
exceeds a Hubble time in most halos of interest (see their Appendix B). In practice, they found that a maximum asymptotic
local velocity dispersion of $\sigma_{i,{\rm max}}(R)=\sigma_{\rm DM}(R)$ provided a sufficiently accurate description of their
numerical results on timescales $\lesssim 10\,{\rm Gyr}$. We adopt this limiting velocity dispersion in what follows.

The finite asymptotic velocity dispersion of stellar particles implies that equation~(\ref{eq:lin}) will break down as
$\sigma_i\rightarrow\sigma_{\rm DM}$. \citet{Ludlow2021} found that a better description of their numerical results was
obtained using
\begin{equation}
  \sigma_i^2=\sigma_{\rm DM}^2\biggr[1-\exp\biggr(-\frac{t+t_0}{t_{\sigma_i}}\biggl)\biggl],
  \label{eq:exp}
\end{equation}
where $t_{\sigma_i}$ is the timescale at which the linear heating model, equation~(\ref{eq:lin}), predicts $\sigma_i=\sigma_{\rm DM}$, i.e.
\begin{equation}
  \frac{t_{\sigma_i}}{t_c}=\biggr[\sqrt{2}\,\pi\,k_i\, \ln\Lambda \biggr(\frac{\rho_{\rm DM}}{\rho_{200}}\biggl)^{\alpha_i}\biggr(\frac{V_{200}}{\sigma_{\rm DM}}\biggl)^2\, \biggl]^{-1},
  \label{eq:tvir}
\end{equation}
where $t_c$ is a characteristic timescale defined by
\begin{equation}
  t_c=\frac{V_{200}^3}{G^2\,\rho_{\rm DM}\,m_{\rm DM}}.
  \label{eq:tc}
\end{equation}
The timescale $t_0$ in equation~(\ref{eq:exp}) is defined such that $\sigma_i(t_0)=\sigma_{i,0}$ is the initial velocity dispersion in
the $i$ direction, and may be calculated using
\begin{equation}
  \frac{t_0}{t_{\sigma_i}}=\ln\biggr(\frac{\sigma_{\rm DM}^2}{\sigma_{\rm DM}^2-\sigma_{i,0}^2}\biggl).
  \label{eq:t0}
\end{equation}
It is easy to verify that equation~(\ref{eq:exp}) reduces to equation~(\ref{eq:lin}) in the limits $\sigma_{i,0}\rightarrow 0$
and $\sigma_i\ll \sigma_{\rm DM}$.
As we discuss below (and show in more detail in \Cref{sec:fitting} and in \citealt{Ludlow2021}), equation~(\ref{eq:exp}) provides
an accurate description of the
velocity dispersion profiles of galactic stellar discs and how they evolve in response to spurious collisional heating. For
example, the dashed lines in Fig.~\ref{fig:kinematic-profiles} show the best-fit radial $\sigma_\phi$ profiles (upper-left panel)
and the time evolution of $\sigma_\phi$ (measured at $R_{1/2}$; upper-right panel) obtained using equation~(\ref{eq:exp}).

We are also interested in the evolution of the mean rotation velocity of the disc ($\overline{v}_\phi$). This is closely related
to the evolution of the velocity dispersion ($\sigma_\phi$) -- the well known phenomenon of asymmetric drift. The dynamical
relation between $\overline{v}_\phi$ and $\sigma_\phi$ (or $\sigma_R$) is usually truncated at leading order in $\sigma_\phi/V_c$
(or $\sigma_R/V_c$),
a good approximation for the observed stellar motions in the solar neighbourhood but inadequate for simulated discs with high levels
of spurious heating. In this case, we must appeal to a more general form of the Jeans equation \citep[e.g.][\textsection 4.8.2]{Binney2008}.
For an axi-symmetric disc with an exponential surface density profile (i.e. equation~\ref{eq:star-density}) that is symmetric about the
$z$-axis and has no bulk radial motions (i.e. $\overline{v}_z=\overline{v}_R=0$), this can be written as
\begin{equation}
  \overline{v}_\phi^2=V_c^2-\sigma_\phi^2+\sigma_R^2\biggr(1-\frac{R}{R_d}\biggl)+R\,\frac{\partial\sigma_R^2}{\partial R}.
  \label{eq:asymm_drift}
\end{equation}
Note that equation~(\ref{eq:asymm_drift}) applies to a steady state disc and also assumes that orbits in the vertical and radial
directions are decoupled (i.e. $\overline{v_R\, v_z}\approx 0$), but does {\em not} assume $\sigma_i\ll \sigma_{\rm DM}$.

The dotted lines in the lower-left panel of Fig.~\ref{fig:kinematic-profiles} show the radial $\overline{v}_\phi$ profiles
expected from equation~(\ref{eq:asymm_drift}) assuming $\sigma_R$ and $\sigma_\phi$ are given by the best-fit
equation~(\ref{eq:exp}), and that the disc scale-length is equal to its initial value (for clarity, these lines are
    plotted over the same radial range resolved by each simulation).
    These curves reproduce our numerical results reasonably well, at least in the outer parts of discs. In the inner parts, however,
    differences are evident, which are mainly due to the evolving surface density profile of the stellar disc, which complicates the
    application of equation~(\ref{eq:asymm_drift}). In fact, we find better agreement between our simulated results and
    equation~(\ref{eq:asymm_drift}) when using the {\em evolving} disc scale length, but since we cannot predict the latter we instead
    opt for a simpler approach to model $\overline{v}_\phi(R)$.

Specifically, we find that the $\overline{v}_\phi(R)$ profiles are also accurately described by set of equations
analogous to the ones used for $\sigma_i$. One such equation can be written as
\begin{equation}
  \overline{v}_\phi=V_c\,\exp\biggr(-\frac{t+t_0}{t_{v_\phi}}\biggl),
  \label{eq:exp_phi}
\end{equation}
where $V_c$ includes contributions from the disc and halo, and $t_{v_\phi}$ is a characteristic timescale
over which $\overline{v}_\phi$ is reduced relative to its initial value (i.e. $\overline{v}_{\phi,0}$).
We determine $t_{v_\phi}$ by matching the slopes of equation~(\ref{eq:exp}; for $\sigma_\phi$) and equation~(\ref{eq:exp_phi})
at $t=0$ for an initially thin disc (i.e. $v_{\phi,0}\approx V_c$ and $\sigma_{i,0}\approx 0$), which implies
\begin{equation}
  \frac{t_{v_\phi}}{t_{\sigma_\phi}}=-\frac{d(\sigma_\phi^2)}{d v_\phi}\biggr(\frac{V_c}{\sigma_{\rm DM}^2}\biggl).
  \label{eq:tvir_phi}
\end{equation}
We identify the first term on the right-hand side of equation~(\ref{eq:tvir_phi}) with the ratio of the second- and
first-order moments of the parallel velocity change $\Delta v_{||}$ of a star particle relative to a background
population of DM particles, which were derived by \citet{Chandrasekhar1960} and applied to disc galaxies by \citet{Lacey1985}
using epicyclic theory. Assuming a relative velocity $v_{\rm rel}$ between
stellar and DM particles, this yields
\begin{equation}
  \frac{t_{v_\phi}}{t_{\sigma_\phi}}=\frac{2\,V_c}{v_{\rm rel}}\equiv f,
  \label{eq:tvir_phi2}
\end{equation}
where $f$ is a number of order unity that we treat as a free parameter. The value of $t_0$ in equation~(\ref{eq:exp_phi}) is
defined such that $\overline{v}_\phi(t=0)=v_{\phi,0}$ is the initial azimuthal velocity of the disc, i.e.
\begin{equation}
  \frac{t_0}{t_{v_\phi}}=\ln\biggr(\frac{V_c}{\overline{v}_{\phi,0}}\biggl).
  \label{eq:t0-velocity}
\end{equation}

Although the asymmetric drift equation provides a reasonable approximation to the $\overline{v}_\phi$ profiles obtained from
our simulations, we prefer to use equation~(\ref{eq:exp_phi}) because:
1) it predicts $\overline{v}_\phi\rightarrow 0$ for $t\gg t_{v_{\phi}}$ (equation~\ref{eq:asymm_drift}
does not); and 2) it can be calculated from properties of the galaxy's DM halo, with no reference to the structure of the galaxy itself.
We expect the latter point to be beneficial when applying our model to galaxies and DM haloes identified in cosmological
simulations of galaxy formation, whose structural properties may or may not have been affected by spurious collisional heating.

To obtain the best-fit values of $\alpha_i$ and $k_i\, \ln\Lambda$ (for $\sigma_i$), and $f$ (for $\overline{v}_\phi$)
we fit equations~(\ref{eq:exp}) and (\ref{eq:exp_phi}) to the measured azimuthal velocity dispersions and mean azimuthal
velocities, respectively, obtained from a subset of our disc galaxy simulations.
In practice, we combine results from our fiducial runs (i.e. $V_{200}=200\,{\rm km/s}$, $\mu=5$ and $c=10$)
with those obtained from models with higher ($c=15$) and lower ($c=7$) concentration DM halos (which
also use $V_{200}=200\,{\rm km/s}$ and $\mu=5$). We also combine measurements made at a few characteristic radii -- specifically,
$R_{1/4}$, $R_{1/2}$ and $R_{3/4}$ enclosing one quarter, one half and three quarters of all stellar particles, respectively.
Doing so allows us to probe a
larger range of DM densities and velocity dispersions than would be possible by fitting our fiducial models alone. We also
limit our fits to timescales over which $\Delta\sigma_i^2/\Delta t$ and $\Delta v_\phi/\Delta t$ evolve approximately linearly with time
(which roughly corresponds to excluding outputs for which $\sigma_i\gtrsim 0.2\, \sigma_{\rm DM}$).
The best-fit parameters are listed in \Cref{table:best-fit-values} for $\sigma_\phi$ and $\overline{v}_\phi$,
along with updated parameters\footnote{We do not present results in this paper for $\sigma_z$ and $\sigma_R$, as they they
were discussed at length by \citet{Ludlow2021}. We refer to that paper for a detailed discussion of these velocity components.} for
$\sigma_z$ and $\sigma_R$. In \Cref{sec:fitting} we show that our best-fit model also describes the spurious disc heating rates
obtained for galaxies not included in our fits, specifically those corresponding to haloes of different virial mass, different $\mu$
values, as well as discs with different initial scale heights.

The various panels of Fig.~\ref{fig:kinematic-profiles} confirm that our simple empirical model (i.e. equations~\ref{eq:exp} and \ref{eq:exp_phi})
describes the spurious evolution of $\sigma_\phi$ and $\overline{v}_\phi$ reasonably well. The coloured dashed lines
in the left-hand panels, for example, plot the predicted velocity dispersion profiles (upper left) and azimuthal
velocities (lower left) after $t=5\,{\rm Gyr}$. These curves, obtained by extrapolating the corresponding initial
profiles of $\sigma_\phi$ and $\overline{v}_\phi$ (shown as black dot-dashed lines in the corresponding panel),
accurately describe the simulated data at all resolved radii.  In the right-hand panels, dashed coloured lines plot the corresponding
quantities as a function of time, but at the initial half-mass radius of the disc. These predictions agree well with our
numerical results over the entire $9.8\,{\rm Gyr}$ simulated. 

Equation~(\ref{eq:exp}) for $\sigma_i$ and equation~(\ref{eq:exp_phi}) for $\overline{v}_\phi$ are also sufficient to describe the
evolution of several morphology diagnostics presented in \Cref{sec:heating}, provided the latter are evaluated at fixed radii.
Indeed, the dashed lines plotted in the upper panel of Fig.~\ref{fig:predicted-kinematics} show the predicted values of
$\overline{v}_\phi/\sigma_{\rm 1D}$, and in the lower panel the predicted $\lambda_r$ (i.e. equation~\ref{eq:lambda-data}, but using the
predicted velocities rather than the measured ones). For $\kappa_{\rm rot}$ (middle panels) we plot 
\begin{equation}
  \kappa_{\rm rot}=\frac{\overline{v}_\phi^2+\sigma_\phi^2}{\overline{v}_\phi^2+\sum \sigma_i^2},
  \label{eq:kap_mod}
\end{equation}
which is equivalent to equation~(\ref{eq:kappa-data}) in the limit of large particle numbers provided $\overline{v}_z\approx \overline{v}_R\approx 0$,
which is approximately valid for our simulations. Note that an isotropic stellar velocity distribution corresponds to $\kappa_{\rm rot}=1/3$, which
explains why the blue curve in the middle panel of Fig.~\ref{fig:predicted-kinematics} asymptotes to $1-\kappa_{\rm rot}\approx 2/3$. 
Our model reproduces the evolution of all three morphology diagnostics reasonably well. 

Finally, the spheroid-to-total ratio -- $S/T$, defined as twice the mass fraction of counter-rotating orbits -- 
can also be predicted, provided the azimuthal velocity {\em distribution} is known. In practice, we find that $v_\phi$ is approximately
normally distributed, with a mean and standard deviation given by equations~(\ref{eq:exp_phi}) and (\ref{eq:exp}), respectively.
Hence, we may use the approximation
\begin{align}
  S/T &\approx \frac{2}{\sqrt{2\,\pi}\sigma_\phi} \int_{-\infty}^0 \exp\biggr({\frac{-(v_\phi-\overline{v}_\phi)^2}{2\,\sigma_\phi^2}}\biggl)\, dv_\phi,\label{eq:ST1} \\
  &=1+\erf\biggr(-\frac{\overline{v}_\phi}{\sqrt{2}\,\sigma_\phi}\biggr)\label{eq:ST2},    
\end{align}
where $\erf (x)$ is the error function. The various dashed lines in the lower panel of Fig.~\ref{fig:unpredicted-kinematics} show
the evolution of the spheroid-to-total ratios of our discs as anticipated by equation~(\ref{eq:ST2}). We note that
predicting the disc-to-total mass ratio is more challenging, as it requires an accurate model of the distribution of
$\varepsilon_{\rm circ}$ and how it evolves due to spurious collisional heating.
We defer a more detailed analysis of the evolution of stellar particle orbital circularities to future work.

\begin{table}
  \caption{Best-fit values for the free parameters $k_i\, \ln\Lambda$ and $\alpha_i$
    obtained from equation~(\ref{eq:exp}) (for $\sigma_\phi$, $\sigma_z$, and $\sigma_R$) and the best-fit value of $f$
    obtained from equation~(\ref{eq:exp_phi}) (for $\overline{v}_\phi$) for the different velocity components in our model.}
  \label{table:best-fit-values}
  \centering
  \begin{tabular}{c c r r r}
    \hline \hline
    Vel. Component & eq. & $k_i \, \ln \Lambda$ & $\alpha_i$ & $f$ \\
    \hline
    $\sigma_z$          & eq.~(\ref{eq:exp})     & 20.19 & -0.308 & $-$ \\
    $\sigma_R$          & eq.~(\ref{eq:exp})     & 20.17 & -0.189 & $-$ \\
    $\sigma_\phi$       & eq.~(\ref{eq:exp})     &  9.40 & -0.115 & $-$ \\
    $\overline{v}_\phi$ & eq.~(\ref{eq:exp_phi}) & $-$   & $-$    & 0.75 \\
    \hline
  \end{tabular}
\end{table}

\subsection{Implications for disc galaxies in cosmological simulations}
\label{ssec:cosmosims}

\begin{figure*}
  \includegraphics[width=0.9\textwidth]{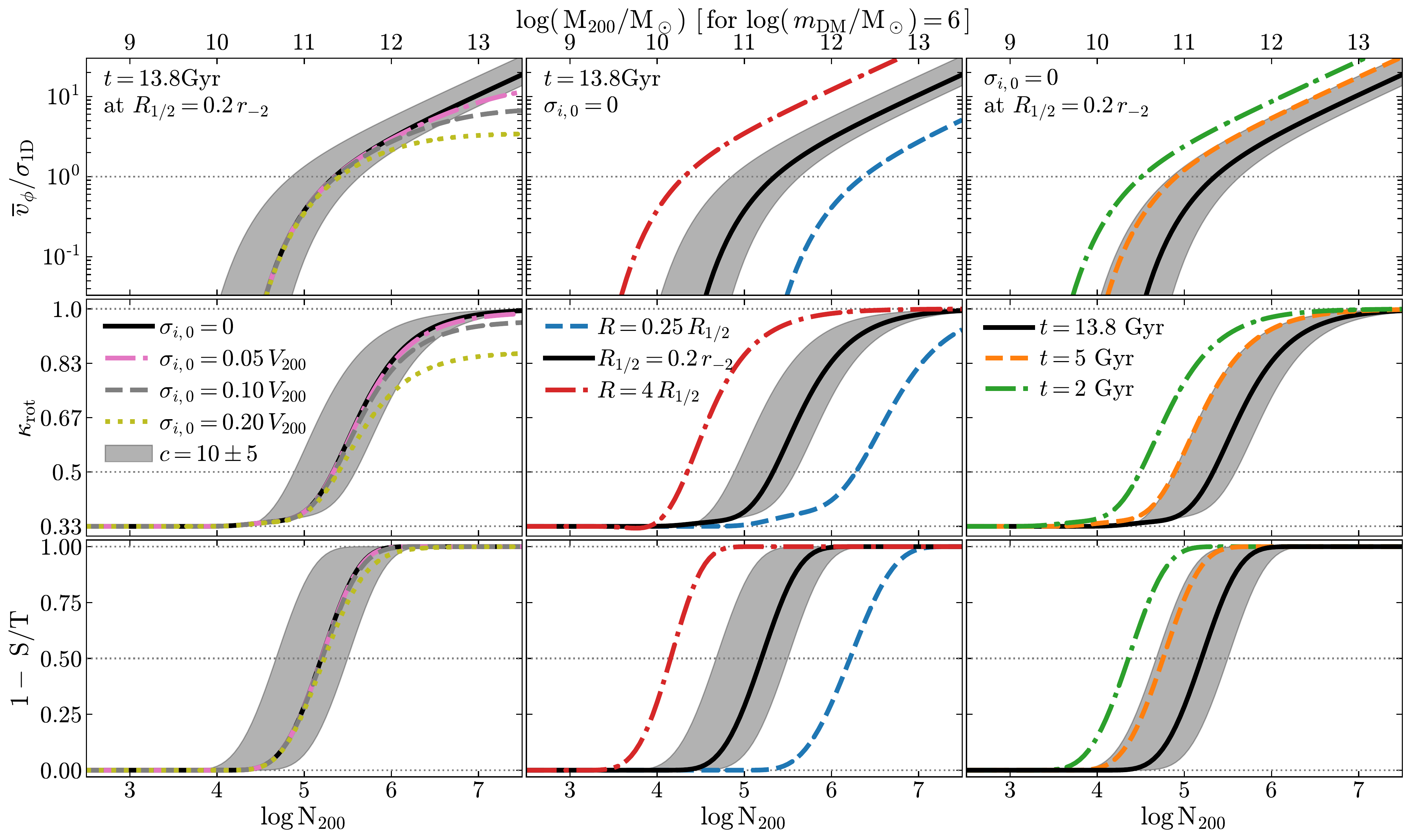}
  \caption{Model predictions for the spurious evolution of galactic disc morphology plotted as a function of the number of DM halo
    particles, ${\rm N_{200}}$, within the virial radius, $r_{200}$.
    From top-to-bottom, we plot the ratio of rotation-to-dispersion velocities, $\overline{v}_\phi/\sigma_{\rm 1D}$, the
    rotation parameter, $\kappa_{\rm rot}$ (i.e. equation~\ref{eq:kap_mod}), and the spheroid-to-total mass ratio, $1-S/T$ (equation~\ref{eq:ST2}).
    Left-hand panels highlight the impact of
    different initial disc velocity dispersions, and correspond to the predicted morphology at the disc characteristic scale
    radius, $R=R_{1/2}$, after $t=13.8\,{\rm Gyr}$ of evolution
    (note: we assume that initially $\sigma_{i,0}=\sigma_{\phi,0}=\sigma_{z,0}=\sigma_{R,0}$).
    The middle panels show results for an initially {\em cold} disc (i.e. $\sigma_{i,0}=0$) but at different
    galacto-centric radii; specifically, $R_{1/2}$ (solid black line), $R_{1/2}/4$ (dashed blue line) and $4\, R_{1/2}$ (dot-dashed red line).
    The right-hand panels show results at $R_{1/2}$ for initially cold discs of different age: $t=2\,{\rm Gyr}$ (dot-dashed green line),
    $5\,{\rm Gyr}$ (dashed orange line), and $13.8\, {\rm Gyr}$ (solid black line). In all cases we model the DM halo using an NFW profile with
    concentration parameter $c=10$; the grey shaded regions show the effect of varying $c$ between 5 and 15, but for clarity
    these results are only shown for the subset of models with $\sigma_{i,0}=0$, $t=13.8\,{\rm Gyr}$, and for $R=R_{1/2}$.
    In all cases, we assume that $R_{1/2}=0.2\times r_{-2}$, where $r_{-2}=r_{200}/c$ is the characteristic scale radius of the
    NFW halo, which closely traces the scale radii of observed discs \citep[e.g.][]{Ferrero2017}. For comparison, we annotate the upper
    axes with values of ${\rm M_{200}}$ corresponding to a DM particle mass of $m_{\rm DM}=10^6\, {\rm M_\odot}$.}
  \label{fig:n-dependence}
\end{figure*}

The spurious collisional heating of stellar motions in galactic discs may have consequences for the evolution of galaxy morphologies in
cosmological simulations. In this section, we use the best-fit model described above to make predictions for several
morphology diagnostics, as well as how they depend on galaxy age, DM halo structure, and distance from the galaxy centre.

A few results are plotted as a function of the number of DM particles per halo, ${\rm N_{200}}$, in the various
panels of Fig.~\ref{fig:n-dependence}. We focus on three morphology indicators: $\overline{v}_\phi/\sigma_{\rm 1D}$ (top panels),
$\kappa_{\rm rot}$ (middle panels), and $1-S/T$ (bottom panels). When implementing our model, we assume an NFW profile for the DM halo,
and that the half mass radius of the disc can be approximated as $R_{1/2}=0.2\times r_{-2}$,
where $r_{-2}=r_{200}/c$ is the halo's characteristic radius \citep[see][]{Navarro2017, Ferrero2017}.
For haloes with spin parameter $\lambda_{\rm DM}=0.03$ and concentration $c = 10$, this radius corresponds to an angular momentum
retention fraction of $f_j \approx 1.0$. For simplicity, we also assume a massless disc, in which case
the circular velocity term in equation~(\ref{eq:exp_phi}) is due solely to the DM halo with no contribution from the disc.

In the left-hand panels of Fig.~\ref{fig:n-dependence}, we plot our model predictions at $R=R_{1/2}$ and after $t=13.8\,{\rm Gyr}$
for discs with different initial velocity dispersions, $\sigma_{i,0}$\footnote{For simplicity we assume the galaxy is initially an isotropic rotator with
  $\sigma_{i,0}=\sigma_{\phi,0}=\sigma_{z,0}=\sigma_{R,0}$, and $\overline{v}_\phi= V_{200} - \sigma_{i,0}$, which is approximately
valid for our fiducial simulations.}; the middle panels plot
results for $\sigma_{i,0}=0$ and $t=13.8\,{\rm Gyr}$ but at different galacto-centric radii (specifically,
at $R_{1/2}/4$, $R_{1/2}$ and $4\, R_{1/2}$); and in the right-hand panels we plot results at $R=R_{1/2}$ but for galaxies with different
ages: $t=2$, $5$ and $13.8\,{\rm Gyr}$ (note that $\sigma_{i,0}=0$ in these cases).  All model curves assume
an NFW halo with concentration parameter $c=10$; the grey shaded regions surrounding the solid black curves highlight the impact of
varying the halo's concentration parameter between $c=5$ and 15 (for clarity, this is only shown for the subset of models with
$\sigma_{i,0}=0$, $t=13.8\,{\rm Gyr}$ and at $R=R_{1/2}$).

The left-hand panels of Fig.~\ref{fig:n-dependence} suggest that halos resolved with fewer than ${\rm a\,\rm few}\times 10^{5}$
DM particles are vulnerable to strong spurious morphological evolution after a Hubble time, regardless of the disc's initial velocity
dispersion. For illustration, consider the results obtained for $c=10$ and $\sigma_i=0$, which are shown as solid black lines in each
of the left-hand panels. For ${\rm N_{200}=10^{5.5}}$, the value of $\overline{v}_\phi/\sigma_{\rm 1D}$, although initially infinite, drops
to $\approx 1.36$ after a Hubble time. Similarly, $\kappa_{\rm rot}$ drops from 1 to $\approx 0.61$ for the same ${\rm N_{200}}$.
The spheroid-to-total ratio, however, appears to be a more robust measure of morphology, at least when measured at $R=R_{1/2}$. Indeed,
we find $S/T \approx 0.18$ after $t=13.8\,{\rm Gyr}$ for ${\rm N_{200}}=10^{5.5}$. 
For ${\rm N_{200}}=10^{4.5}$, however, our model predicts that spurious collisional heating will have catastrophic consequences for
simulated discs. In this case, $\overline{v}_\phi/\sigma_{\rm 1D}\approx 0.02$, $\kappa_{\rm rot}\approx 0.34$, and $S/T \approx 0.98$,
after a Hubble time. Note too that the values quoted here differ only slightly
for initially hotter discs (the various curves in the left-hand panels approximately overlap).

The middle panels show, unsurprisingly, that collisional heating drives more substantial morphological evolution near galaxy
centres than in their outskirts. This is due to the strong radial gradient in the number density of DM particles across the
disc (which also explains why galaxies in higher concentration haloes are more strongly affected than those in lower concentration ones).
As a result, larger particle numbers are required to suppress spurious morphological evolution in the central regions of galaxies
than in their outer parts. For example, after $t=13.8\,{\rm Gyr}$ we find that maintaining $\overline{v}_\phi/\sigma_{\rm 1D}\gtrsim 4$
requires ${\rm N_{200}}\gtrsim 2\times 10^7$ at $R=R_{1/2}/4$, but only ${\rm N_{200}}=1.3\times 10^5$ at $4\,R_{1/2}$ (at $R=R_{1/2}$, it requires
${\rm N_{200}=1.6\times 10^6}$).

Interestingly, this implies that spurious collisional heating drives in inside-out evolution of disc morphology, affecting first
small radii, and only later affecting the outer disc structure. The
bottom-middle panels of Fig.~\ref{fig:n-dependence} imply that, at least in the simplest
case of an initially cold disc, spurious heating will first transform the central regions into a dispersion supported structure, leaving
the outer disc largely unaffected. This is indeed what our model predicts for DM haloes with ${\rm N_{200}}\approx 10^5$ (and
$c=10$): in this case, after $t=13.8\,{\rm Gyr}$, we find $S/T \approx 0$ at $R=4\, R_{1/2}$, but $S/T \approx 1$ at $R=R_{1/2}/4$.

Of course, the stellar components of galaxies are typically not contemporaneous, but rather formed smoothly over time or through
episodic bursts of star formation. Although these complexities are ignored in our model, the rightmost panels of Fig.~\ref{fig:n-dependence}
give an impression of the expected magnitude of the effect. There, the solid black, dashed orange, and dot-dashed green curves plot the
various morphology diagnostics for galaxies or stellar populations of different age -- specifically, $t=13.8\,{\rm Gyr}$,
$5\,{\rm Gyr}$, and $2\,{\rm Gyr}$, respectively. Clearly younger galaxies (or populations of stars) are less affected by spurious
collisional heating, but they are not less vulnerable. Because collisional heating is a cumulative effect, given sufficient time, {\em all}
simulated galaxies will, to some extent, suffer the consequences of their finite resolution.

\section{Discussion and Conclusions}
\label{sec:summary}

We used a suite of idealised simulations of equilibrium stellar discs embedded within ``live'' DM haloes to study the
effects of spurious collisional heating of star particles by DM particles on disc kinematics and morphology. Our simulated discs,
previously presented in \citet[][which we refer to for more details]{Ludlow2021}, are constructed to be in a state of stable
equilibrium, are DM dominated, free from disc instabilities, contain no gaseous component, experience no mergers or accretion of DM
or baryons, and are non-cosmological. Although this presents a highly simplified scenario, it allows us to isolate
the effects of spurious collisional heating on the kinematics of stellar discs, while eliminating other genuine sources of heating.
Our main results can be summarised as follows:

\begin{itemize}
  
\item The cumulative effects of spurious collisional heating alter the
  visual morphology of simulated galactic discs (Figs.~\ref{fig:projections-low} to~\ref{fig:projections-hi}), making them
  appear thicker and more spheroidal than they were initially. Both the scale length and scale height of discs increase as
  a result, but at fixed mass resolution the increase in the latter occurs more rapidly (Fig.~\ref{fig:shape-global}).
  At fixed halo mass, the net effect depends strongly on the mass of DM particles (or equivalently, on the number of DM halo
  particles), but also on the local DM density and velocity dispersion. Our fiducial Milky Way-mass systems, for example, initially
  have an aspect ratio of $z_{1/2}/R_{1/2}\approx 0.02$, but after
  $t=10\,{\rm Gyr}$ have $z_{1/2}/R_{1/2}\approx 0.05$ for $m_{\rm DM}=10^6\,{\rm M_\odot}$ ($N_{\rm DM}=1.8 \times 10^6$) and
  $\approx 0.16$ for $m_{\rm DM}=10^7\,{\rm M_\odot}$ ($N_{\rm DM}=1.8 \times 10^5$). 
  
\item Provided $\sigma_\phi\ll\sigma_{\rm DM}$, the azimuthal velocity dispersion of stellar particles (squared) increases
  approximately linearly with time, i.e. $\sigma_\phi^2\propto t$, as predicted by the analytic model of \citet{Lacey1985}. In poorly-resolved
  systems, however, the azimuthal velocity dispersion reaches a maximum asymptotic value set by the local velocity dispersion
  of the DM halo, i.e. $\sigma_{\phi,{\rm max}}\approx \sigma_{\rm DM}$ (a similar result holds for the vertical and radial
  velocity dispersion of stellar particles; see \citealt{Ludlow2021} for details), and is better described by equation~(\ref{eq:exp}).
  We also found that the evolution of the mean rotation velocity of the disc is accurately described by asymmetric drift, although
      doing so requires knowledge of how its surface density profile evolves due to spurious collisional heating, which is
      difficult to predict. For that reason, we provided a
  parametric model (equation~\ref{eq:exp_phi}; similar to equation~\ref{eq:exp} for $\sigma_i$) that accurately describes the
  evolution of the $\overline{v}_\phi(R)$ profiles for all of our simulations, and over all resolved radii.
  Equation~(\ref{eq:exp_phi}) has the added benefit that it depends only on the structure of a galaxy's DM halo, which is better
  resolved and easier to predict than the structure of the galaxy itself, which evolves due to collisional heating.
  We note, however, that the spurious increase in the velocity dispersion of stellar particles is largely
  a result of kinetic energy being transferred from their azimuthal motion into vertical and radial motions, with little
  change to the total kinetic energy of the stellar component. For example, the total kinetic energy of our
  ($m_{\rm DM}=10^7\,{\rm M_\odot}$) fiducial disc increased by only about 5 percent over a Hubble time as a result of collisional
  heating, but the kinetic energy contributed by velocity dispersion increased by a factor of $\approx 5$ over the same time interval.

\item Spurious collisional heating alters the kinematics of disc stars, as well as the sizes and shapes of discs, resulting in
  systematic biases in the location of galaxies in a number of standard galaxy scaling relations, including the
  size-mass relation, the \citet{Tully1977} relation, and the \cite{Fall1983} relation
  (Fig.~\ref{fig:scaling-relations}). Galaxies that are affected by spurious heating should therefore
  be eliminated from simulated galaxy populations before they are compared to observed galaxy samples. Not doing so risks drawing
  erroneous conclusions about galaxy formation theory, or about the virtues or weaknesses of galaxy formation models.

\item All of the galaxy morphology diagnostics we have studied are sensitive to the effects of spurious collisional heating: Discs get
  thicker, have increased levels of dispersion support, and larger axis ratios due to its cumulative effects. Spurious heating also
  broadens the circularity
  distributions of stellar particle orbits, $\varepsilon_{\rm circ}$, resulting in higher spheroid-to-total mass ratios (commonly
  estimated as twice the mass fraction with $\varepsilon_{\rm circ}<0$) and lower disc-to-total mass ratios (often defined
  as the stellar mass fraction with $\varepsilon_{\rm circ}\geq 0.7$). For quantities measured at the galaxy half mass radius, $R_{1/2}$,
  the effects are noticeable for haloes resolved with
  fewer than $\approx 10^6$ DM particles, and we predict that those resolved with fewer than $\approx 10^5$ particles are
  unlikely to harbour old stellar discs at all (Fig.~\ref{fig:n-dependence}). Suppressing spurious morphological evolution at radii
  smaller than $R_{1/2}$ requires haloes to be resolved with even more DM particles. 

\end{itemize}

Our results are however obtained from a particular set of highly idealised simulations, and therefore may not represent
the typical simulated galaxy's response to spurious collisional heating. For example, the disc mass fractions are too low ($f_\star=0.01$, whereas observations
suggest $f_\star\approx 0.05$ for the discs of Milky Way-mass galaxies; e.g. \citealt{Posti2021}) and they contain no gas component. This deliberate
choice was made to suppress the formation of local instabilities that can contribute to disc heating, permitting us to isolate the numerical
effects. But instabilities are believed to
be present in real galaxies, and may also be present in those formed in cosmological simulations provided they have sufficient mass
and spatial resolution. Our discs also contain no giant molecular clouds, globular clusters, or spiral density waves, all of which
provide additional sources of gravitational heating \citep[e.g.][]{Hanninen2002,Aumer2016,Gustafsson2016}. Our runs are non-cosmological and therefore
eliminate the possibility of accretion-driven heating, or tidal heating due to mergers or fly-bys, all of which are likely to heat the stellar
components of real galaxies as well as those in cosmological simulations \citep[e.g.][]{Benson2004,Genel2018,Borrow2022}. 

This suggests that the heating rates of stellar discs in cosmological simulations may in fact be {\em larger} than those inferred from our
idealised runs, because there are contributions from both numerical and physical effects. However, our discs have no intrinsic stellar age
gradients and no star formation,
whereas those in cosmological simulations can in principle form new stars and thereby replenish their thin discs. Star formation will therefore
combat the cumulative effects of collisional heating since new stars are typically born in thin discs with low velocity dispersion
($\lesssim 10\,{\rm km/s}$). 

Thus, it will be difficult to establish the extent to which spurious heating affects the properties of galaxies in cosmological simulations,
and doing so will likely require simulations that specifically aim to suppress the effect. We will present results from such a simulation
in a forthcoming paper.

\section*{Acknowledgements}
MJW acknowledges support from the Australian Government
Research Training Program (RTP) Scholarship. 
ADL and DO acknowledge financial support from the Australian Research Council through their Future Fellowship
scheme (project numbers FT160100250, FT190100083, respectively). 
CL has received funding from the ARC Centre of Excellence for All Sky Astrophysics in 3 Dimensions (ASTRO 3D), through project number CE170100013.
This work has benefited from the following public
{\textsc{python}} packages: {\textsc{scipy}} \citep{scipy}, {\textsc{numpy}} \citep{numpy}, {\textsc{matplotlib}}
\citep{matplotlib} and {\textsc{ipython}} \citep{ipython}.

\section*{Data Availability}

Our simulation data can be made available upon request or can be
generated using publicly available codes. Our model results can be
reproduced using the equations provided in the paper.


\bibliographystyle{mnras}
\bibliography{mybib} 


\appendix

\section{Comparison of the disc heating model to a diverse set of disc galaxy simulations}
\label{sec:fitting}

\begin{figure*}
  \includegraphics[width=0.9\textwidth]{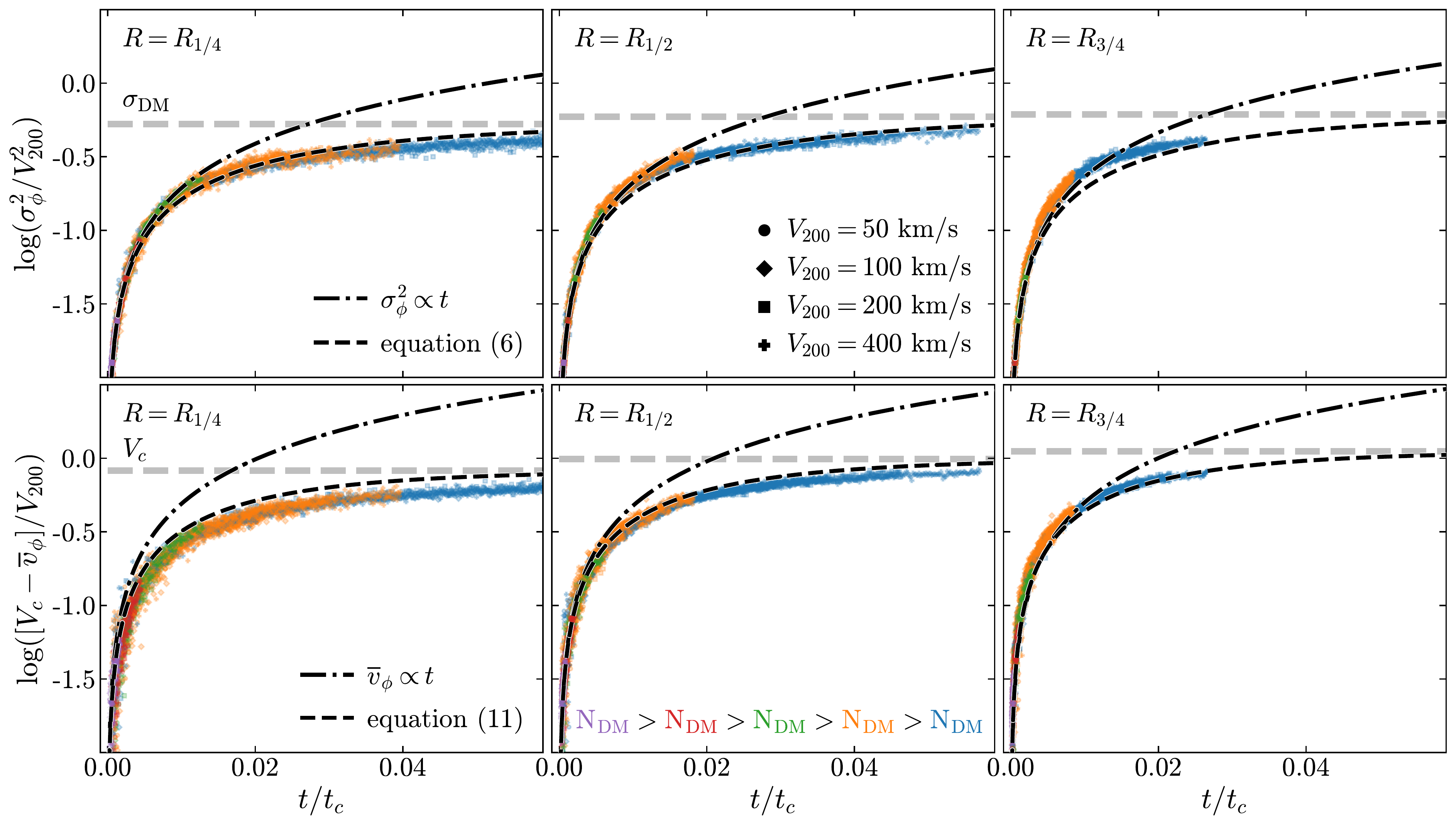}
  \caption{Evolution of the azimuthal velocity dispersion (upper panels) and mean azimuthal velocity (lower panels), normalised by
    $V_{200}$, as a function of (dimensionless) time, $t/t_c$. Results are shown for our fiducial runs, i.e. $V_{200}=200\,{\rm km/s}$,
    as well as for $V_{200}=50$, 100, and $400\,{\rm km/s}$ (different symbols);
    all runs have $\mu=5$, $\lambda_{\rm DM}=0.03$, $f_\star=0.01$, and $c=10$.
    From left to right, different columns plot results measured at the characteristic radii enclosing one quarter (i.e. $R_{1/4}$),
    one half (i.e. $R_{1/2}$) and three quarters (i.e. $R_{3/4}$) of the galaxy's initial stellar mass, respectively. For each $V_{200}$,
    we use different colours to distinguish runs carried out with different $m_{\rm DM}$. At early times, the evolution of both
    $\overline{v}_\phi$ and $\sigma_\phi$ can be approximated reasonably well by a power-law,
    as shown by the dot-dashed black lines. At late times, however, substantial departures from power-law heating rates are evident as
    galaxies approach their maximally heated states, which corresponds to $\sigma_{\phi, \rm max} = \sigma_{\rm DM}$ (dashed lines,
    upper panels) and $\overline{v}_{\phi, \rm max} = 0$ (dashed lines, lower panels). 
    In this regime, $\sigma_\phi$ and $\overline{v}_\phi$ are better described by
    equations~(\ref{eq:exp}) and (\ref{eq:exp_phi}), respectively, which are plotted as dashed black lines in each panel. }
  \label{fig:A1}
\end{figure*}

\begin{figure*}
  \includegraphics[width=0.9\textwidth]{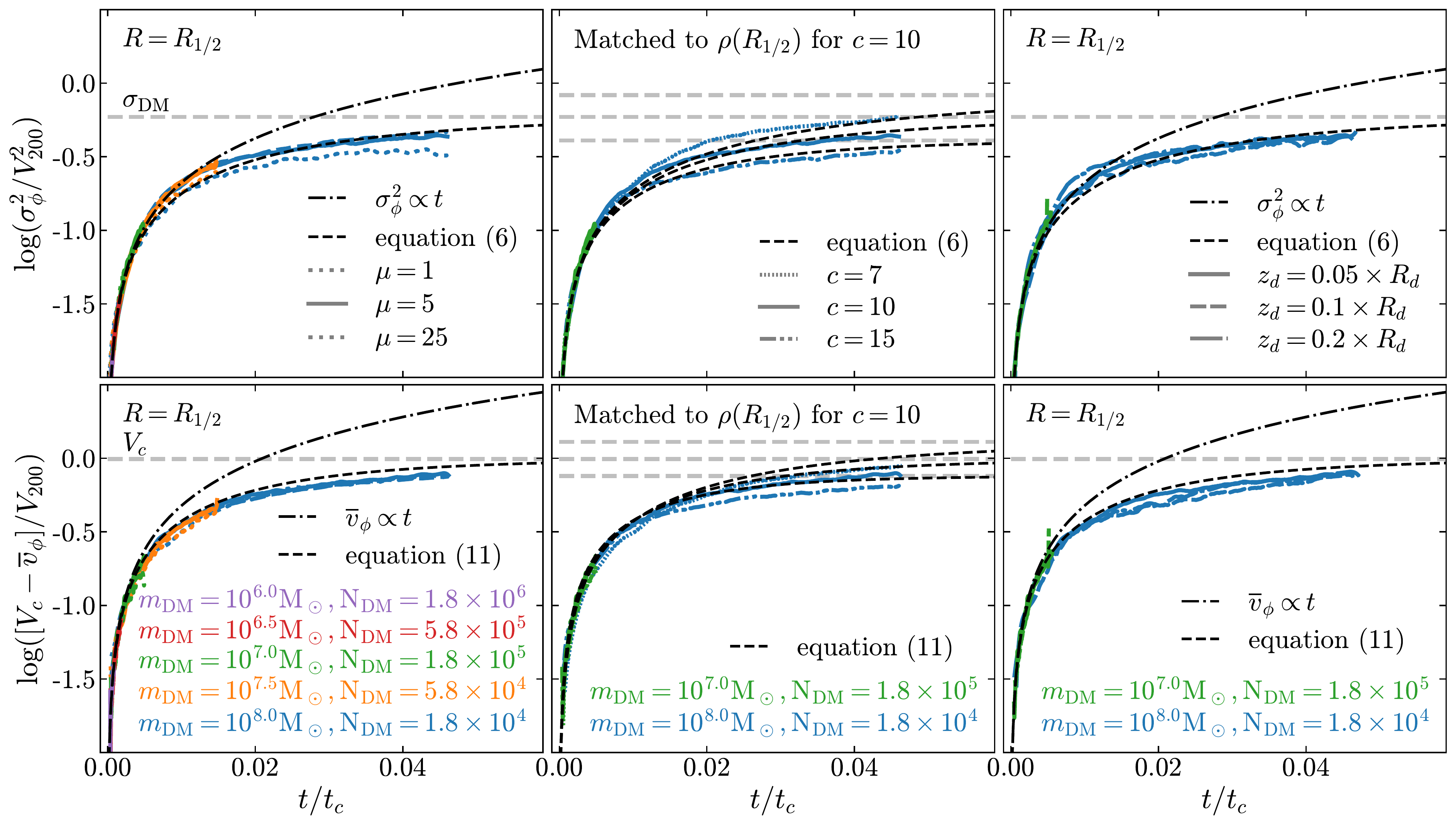}
  \caption{Same as Fig.~\ref{fig:A1}, but for models with different DM-to-stellar particle mass ratios, $\mu$ (left-hand panels),
    different DM halo concentrations (middle panels), and different initial disc thickness (right-hand panels). Properties of the
    discs and haloes that are not varied are held fixed to the fiducial values, i.e. $V_{200}=200\,{\rm km/s}$, $f_{\star}=0.01$,
    $\lambda_{\rm DM}=0.03$, $c=10$ and $z_d=0.05\, R_d$. All velocities are measured at the initial half-mass radius of the disc,
    except those in the middle panels corresponding to haloes with $c=7$ and $15$. In those cases, velocities are measured at the
    radii corresponding to the same local DM density as our fiducial ($c=10$) run.}
  \label{fig:A2}
\end{figure*}

The results presented in the main body of our paper were obtained from a set of ``fiducial'' disc galaxy simulations.
These models all adopted the same structural properties for the disc and halo -- namely, $V_{200}=200\,{\rm km/s}$, $f_\star=0.01$,
$\lambda_{\rm DM}=0.03$, $c=10$, $f_j=1$, and $\mu=5$ -- but used different dark matter and stellar particle masses to assess
the impact of spurious collisional heating on the disc. Below we present the azimuthal velocity
evolution for models that vary some of the relevant properties of the disc or halo, while holding others fixed.
Because the various morphology diagnostics we considered in Section~\ref{ssec:shapes} can be calculated from
$\sigma_\phi$ and $\overline{v}_\phi$, we do not consider them explicitly below.

In Fig.~\ref{fig:A1}, we plot the $\sigma_\phi$ (upper panels) and $\overline{v}_\phi$ (lower panels) evolution for haloes of different
virial mass, corresponding to $V_{200}=50\,{\rm km/s}$, $100\,{\rm km/s}$ and $400\,{\rm km/s}$ (in addition to our fiducial runs, i.e. $V_{200}=200\,{\rm km/s}$),
with all other dimensionless properties of the disc and halo held fixed. From left-to-right, the columns correspond to measurements made
at three separate characteristic radii, specifically $R_{1/4}$, $R_{1/2}$, and $R_{3/4}$, enclosing one-quarter, one-half, and three-quarters
of each galaxy's initial stellar mass. Different colour curves correspond to the different particle
masses listed in Table~\ref{table:simulation-list}, which differ for the different values of $V_{200}$. Note that the velocities have been normalised by $V_{200}$
and times by the characteristic timescale $t_c$ (equation~\ref{eq:tc}). In these dimensionless units, the $\sigma_\phi$ and
$\overline{v}_\phi$ evolution obtained for all models evolve similarly regardless of $V_{200}$ or DM particle mass, and all
are accurately described by equations~(\ref{eq:exp}) and (\ref{eq:exp_phi}), respectively.

In Fig.~\ref{fig:A2} we plot the evolution of $\sigma_i$ and $\overline{v}_\phi$ (measured at the initial value of $R_{1/2}$, and normalised
by $V_{200}$) for a suite of models with $V_{200}=200\,{\rm km/s}$ that vary the DM-to-stellar particle mass ratio, $\mu$ (left-hand panels),
the concentration of the DM halo (middle panels), and the initial scale height of the disc, $z_d$ (right-hand panels). As in Fig.~\ref{fig:A1},
different coloured lines correspond to different DM particle masses, as indicated in the legends. 
In agreement with \citet{Ludlow2021}, we find that the effects of spurious collisional heating are largely independent of
$\mu$, at least for the range of particle mass ratios ($1\leq\mu\leq 25$) and timescales ($\lesssim 10\,{\rm Gyr}$) we have
considered in our analysis.

In the middle panels of Fig.~\ref{fig:A2} we verify that our model also describes reasonably well the rate of disc heating in
haloes with different concentrations of DM (note that for $c=7$ and $15$ we have adjusted $f_j$ so that the stellar mass profile is
fixed for all models). In this case we plot heating rates measured at $R=R_{1/2}$ for our fiducial ($c=10$) models, but at the radii corresponding
to the same local DM density for the haloes with higher and lower concentration values. By doing so, the initial heating rates -- which are
dominated by $\rho_{\rm DM}$ -- are approximately equal for all models, but the asymptotic values of $\sigma_\phi$ and $\overline{v}_\phi$
are not.

Finally, in the right-hand panels of Fig.~(\ref{fig:A2}), we consider discs with varying initial scale heights (or equivalently, different
vertical velocity dispersion). Solid lines show our fiducial models (i.e. $z_d=0.05\,R_d$); discs that are initially two and four
times thicker are shown using dashed and dotted lines, respectively. Despite the small differences in the initial velocity dispersion
of the discs, they rapidly evolve to have a similar kinematic structure that is dominated by the cumulative effects of collisional heating
rather than by the small differences in their initial kinematics.

\bsp	
\label{lastpage}
\end{document}